\documentclass[hyper,12pt,letterpaper]{JHEP3}
\usepackage[centertags]{amsmath}
\usepackage{amssymb,multirow,array}
\usepackage{graphicx}

\newcommand{\prn}[1]{\left ( #1 \right )}
\newcommand{\brk}[1]{\left [ #1 \right ]}

\newcommand{\ben}{\begin{eqnarray}\displaystyle}
\newcommand{\een}{\end{eqnarray}}
\newcommand{\be}{\begin{equation}}
\newcommand{\ee}{\end{equation}}

\newcommand{\can}{{\cal C}_{anom} }
\newcommand{\nn}{\nonumber}

\numberwithin{equation}{section}

\title{Constraints on Anomalous Fluid in Arbitrary Dimensions}
\author{Nabamita Banerjee$^{a}$, Suvankar Dutta$^b$,
Sachin Jain$^c$ , R. Loganayagam$^d$ and Tarun Sharma$^c$\\
$^a$ NIKHEF, Science Park 105,
Amsterdam, The Netherlands\\
$^b$ Department of Physics, \\
Indian Institute of Science Education and Research(IISER)
Bhopal, India\\
$^c$Dept. of Theoretical Physics, \\
Tata Institute of Fundamental Research, \\
Homi Bhabha Rd,Mumbai 400005, India. \\
$^d$Junior Fellow, Harvard Society of Fellows, \\
Harvard University, Cambridge, MA 02138 .\\ 
Email:\ \ {\bf nbaner@nikhef.nl, suvankar@iiserbhopal.ac.in,
sachin@theory.tifr.res.in, nayagam@physics.harvard.edu,
tarun@theory.tifr.res.in
}}

\abstract{Using the techniques developed in arxiv: 1203.3544 we compute the
universal part of the equilibrium partition function
characteristic of a theory with multiple abelian $U(1)$ anomalies in
arbitrary even spacetime dimensions.  This contribution is closely linked to
the universal anomaly induced  transport coefficients  in
hydrodynamics which have been studied before using entropy techniques. Equilibrium partition
function provides an
alternate and a microscopically more transparent way to derive the
constraints on these
transport coefficients. We re-derive this way all the known results on
these transport
coefficients including their polynomial structure which has recently
been conjectured to be linked
to the anomaly polynomial of the theory. Further we link the local
description of
anomaly induced transport in terms of a Gibbs current to the more
global description
in terms of the partition function.}

\preprint{}

\begin{document}

\section{Introduction}\label{Sec:Intro}
Anomalies are a fascinating set of phenomena exhibited by field theories
and string theories. For the sake of clarity let us begin by  
distinguishing between three quite different phenomena bearing that name.

The first phenomenon is when a symmetry of a classical action fails to 
be a symmetry at the quantum level. One very common example of an 
anomaly of this kind is the breakdown of classical scale invariance
of a system when we consider the full quantum theory. This breakdown results
in \textit{renormalization group  flow}, i.e., a scale-dependence 
of physical quantities even in a classically scale-invariant
theory. Often this classical symmetry cannot be restored without
seriously modifying the content of the theory. Anomalies of this
kind are often serve as a cautionary tale to remind us that
the symmetries of a classical action like scale invariance 
will often not survive quantisation.

The second set of phenomena are what are termed as gauge anomalies.
A system is said to exhibit a gauge anomaly if a particular
classical gauge redundancy of the system is no more a 
redundancy at a quantum level. Since such redundancies are often
crucial in eliminating unphysical states in a theory, a gauge
anomaly often signifies a serious mathematical inconsistency
in the theory. Hence this second kind of anomalies serve as
a consistency criteria whereby we discard any theory exhibiting
gauge anomaly as most probably inconsistent.

The third set of phenomena which we would be mainly interested 
in this work is when a genuine symmetry of a quantum theory 
is no more a symmetry when the theory is placed in a non-trivial
background where we turn on sources for various operators in the theory.
This lack of symmetry is reflected in the fact that the path integral
with these sources turned on is no more invariant under the original
symmetry transformations. If the sources are non-trivial 
gauge/gravitational backgrounds (corresponding to the charge/energy-momentum
operators in the theory) the path integral is no more gauge-invariant.
In fact as is well known the gauge transformation of the path-integral 
is highly constrained and the possible transformations
are classified by the Wess-Zumino descent relations\footnote{The Wess-Zumino
descent relations are dealt with in detail in various textbooks\cite{Weinberg:1996kr,
Bertlmann:1996xk,Bastianelli:2006rx} and lecture notes \cite{Harvey:2005it,Bilal:2008qx}.}.

Note that unlike the previous two phenomena here we make no 
reference to any specific classical description or the process of
quantisation and hence this kind of anomalies are well-defined
even in theories with multiple classical descriptions (or theories
with no known classical description). Unlike the first kind of
anomalies the symmetry is simply recovered at the quantum level
by turning off the sources. Unlike the gauge anomalies the third
kind of anomalies do not lead to any inconsistency. In what
follows when we speak of anomaly we will always have in mind this
last kind of anomalies unless specified otherwise.

Anomalies have been studied in detail in the least few decades
and their mathematical structure and phenomenological consequence 
for zero temperature/chemical potential situations are reasonably 
well-understood. However  the anomaly related phenomena 
in finite temperature setups let alone in non-equilibrium states 
are still relatively poorly understood despite their obvious relevance
to fields ranging from solid state physics to cosmology. It is becoming
increasingly evident that there are universal transport processes
which are linked to anomalies present in a system and that study
of anomalies provide a non-perturbative way of classifying these
transport processes say in solid-state physics\cite{2012PhRvB..85d5104R}.

While the presence of transport processes linked to anomalies had been
noticed before in a diversity of systems ranging from free fermions\footnote{It would
be an impossible task to list all the references in 
the last few decades which have discovered (and rediscovered) 
such effects in free/weakly coupled theories in 
various disguises using a diversity of methods . See for example 
\cite{Vilenkin:1978hb} for what is probably the earliest study in $3+1d$. See
\cite{Loganayagam:2012pz} for a recent generalisation to arbitrary dimensions.}
to holographic fluids\footnote{See for example \cite{Erdmenger:2008rm,Banerjee:2008th,
Torabian:2009qk} for some of the initial holographic results.}
a main advance was made in \cite{Son:2009tf}. In that work it was
shown using very general entropy arguments that the $U(1)^3$ anomaly 
coefficient in an arbitrary $3+1d$ relativistic field theory is
linked to a specific transport process in the corresponding hydrodynamics.
This argument has since then been generalised to finite temperature
corrections \cite{Neiman:2010zi, Loganayagam:2011mu} and $U(1)^{n+1}$ anomalies in 
$d=2n$ space time dimensions \cite{Kharzeev:2011ds, Loganayagam:2011mu}.
In particular the author of \cite{Loganayagam:2011mu} identified a 
rich structure to the anomaly-induced transport processes
by writing down an underlying Gibbs-current which captured
these processes in a succinct way. Later in a microscopic
context in ideal Weyl gases, the authors of \cite{Loganayagam:2012pz}
identified this structure as emerging from an adiabatic
flow of chiral states convected in a specific way in
a given fluid flow.

While these entropy arguments are reasonably straightforward they
appear somewhat non-intuitive from a microscopic field theory viewpoint.
It is especially important to have a more microscopic understanding of these
transport processes if one wants to extend the study of anomalies 
far away from equilibrium where one cannot resort to such 
thermodynamic arguments.  So it is crucial to first rephrase
these arguments in a more field theory friendly terms so 
that one may have a better insight 
on how to move far away from equilibrium.

Precisely such a field-theory friendly reformulation in $3+1d$ and 
$1+1d$ was found  recently in the references \cite{Banerjee:2012iz} and
\cite{Jain:2012rh} respectively. Our main aim in this paper is to 
generalise their results to arbitrary even space time dimensions. So
let us begin by repeating the basic physical idea behind this 
reformulation in the next few paragraphs.

Given a particular field theory exhibiting certain anomalies, one begins
by placing that field theory in a time-independent gauge/gravitational
background at finite temperature/chemical potential. We take 
the gauge/gravitational background to be spatially slowly 
varying compared to all other scales in the theory.  Using this
one can imagine integrating out all the heavy modes\footnote{Time-independence
at finite temperature and chemical potential essentially means we are 
doing a Euclidean field theory. Unlike the Lorentzian field theory
(which often has light-hydrodynamic modes) the Euclidean field theory
has very few light modes except probably the Goldstone modes arising out
of spontaneous symmetry breaking. We thank Shiraz Minwalla for emphasising
this point.} in the theory to generate an effective Euler-Heisenberg type effective action 
for the gauge/gravitational background fields at 
finite temperature/chemical potential.

In the next step one expands this effective action in a spatial derivative
expansion and then imposes the constraint that its gauge transformation
be that fixed by the anomaly. This constrains the terms that can appear
in the derivative expansion of the Euler-Heisenberg type effective action.
As is clear from the discussion above, this effective action and 
the corresponding partition function have a clear microscopic 
interpretation in terms of a field-theory path integral and hence
is an appropriate object in terms of which one might try to 
reformulate the anomalous transport coefficients. 

The third step is to  link various terms that appear in 
the partition function to the transport coefficients in the
hydrodynamic equations. The crucial idea in this link is the
realisation that the path integral we described above is essentially
dominated by a time-independent hydrodynamic state (or more
precisely a hydrostatic state ). This means in particular that the 
expectation value of energy/momentum/charge/entropy calculated via
the partition function should match with the  distribution
of these quantities in the corresponding hydrostatic state.

These distributions in turn depend on a subset of transport
coefficients in the hydrodynamic constitutive relations which
determine the hydrostatic state. In this way various terms that
appear in the equilibrium partition function are linked to/constrain the
transport coefficients crucial to hydrostatics. Focusing on just
the terms in the path-integral which leads to the failure 
of gauge invariance we can then identify the universal 
transport coefficients which are linked to the anomalies.
This gives a re derivation of various entropy argument results
in a path-integral language thus opening the possibility that
an argument in a similar spirit with Schwinger-Keldysh path integral 
will give us insight into non-equilibrium anomaly-induced phenomena.

Our main aim in this paper is twofold - first is to carry through in 
arbitrary dimensions this program of equilibrium partition function 
 thus generalising the results of \cite{Banerjee:2012iz,Jain:2012rh} and 
re deriving in a path-integral friendly language 
the results of \cite{Kharzeev:2011ds, Loganayagam:2011mu}.

Our second aim is to clarify the relation between the Gibbs current
studied in \cite{Loganayagam:2011mu,Loganayagam:2012pz} and the 
partition function of \cite{Banerjee:2012iz,Jain:2012rh}. Relating
them requires some care on carefully distinguishing the consistent 
from covariant charge , the final result however is intuitive : the 
negative logarithm of the equilibrium partition function 
(times temperature) is simply obtained by integrating the 
equilibrium Gibbs free energy density (viz. the zeroeth
component of the Gibbs free current) over a spatial hyper surface.
This provides a direct and an intuitive link between the local
description in terms of a Gibbs current vs. the global description
in terms of the partition function.

The plan of the paper is following. We will begin by mainly reviewing known 
results in Section \S\ref{sec:prelim}. First we review the formalism/results of 
\cite{Loganayagam:2011mu} in subsection\S\S\ref{subsec:LogaReview} where 
entropy arguments were used to constrain the anomaly-induced transport processes
a Gibbs-current was written down which captured those processes in a succinct way.  
This is followed by subsection\S\S\ref{subsec:PartitionReview} where we 
briefly review the relevant details of the equilibrium partition function formalism
for fluids as developed in \cite{Banerjee:2012iz}. A recap of the relevant results in 
(3+1) and (1+1) dimensions\cite{Banerjee:2012iz,Jain:2012rh} and a 
comparison with results in this paper are relegated to appendix~\ref{app:oldresult}.

Section \S\ref{sec:2ndimu1} is devoted to the derivation of transport coefficients 
for $2n$ dimensional anomalous fluid using the partition function method. The 
next section\S\ref{sec:entropy} contains construction of entropy current for the fluid 
and the constraints on it coming from partition function. This mirrors similar
discussions in \cite{Banerjee:2012iz,Jain:2012rh}. We then compare these results
to the results of \cite{Loganayagam:2011mu} presented before in 
subsection\S\S\ref{subsec:LogaReview} and find a perfect agreement.

Prodded by this agreement, we proceed in next section\S\ref{sec:IntByParts} to a 
deeper analysis of the relation between the two formalisms. We prove an
intuitive relation whereby the partition function could be directly derived 
from the Gibbs current of \cite{Loganayagam:2011mu} by a simple integration
(after one carefully shifts from the covariant to the consistent charge).

This is followed by section\S\ref{sec:2ndimmul} where  we generalise all 
our results for multiple $U(1)$ charges. We perform a $CPT$ invariance 
analysis of the fluid in section \S\ref{sec:CPT} and this imposes 
constraints on the fluid partition function. We end with conclusion
and discussions in section\S\ref{sec:conclusion}.

Various technical details have been pushed to the appendices for the 
convenience of the reader. After the appendix~\ref{app:oldresult}
on comparison with previous partition function results in (3+1) and
(1+1) dimensions, we have placed an appendix~\ref{app:hydrostatics} detailing
various specifics about the hydrostatic configuration considered in 
\cite{Banerjee:2012iz}. We then have an appendix~\ref{app:variationForms}
where we present the variational formulae to obtain currents from
the partition function  in the language of differential forms.
This is followed by an appendix~\ref{app:formConventions}
on notations and conventions (especially the conventions of wedge product etc.).

\section{Preliminaries}\label{sec:prelim}
In this section we begin by reviewing and generalising various results from \cite{Loganayagam:2011mu} where
constraints on anomaly-induced transport in arbitrary dimensions were derived using 
adiabaticity (i.e., the statement that there is no entropy production associated with
these transport processes). Many of the zero temperature results here were also
independently derived by the authors of \cite{Kharzeev:2011ds}. 

We will then review the construction of equilibrium partition function
(free energy) for fluid in the rest of the section. The technique 
has been well explained in \cite{Banerjee:2012iz} and 
familiar readers can skip this part.

\subsection{Adiabaticity and Anomaly induced transport}\label{subsec:LogaReview}

Hydrodynamics is a low energy (or long wavelength) description of a
quantum field theory around its thermodynamic equilibrium. Since the fluctuations are
of low energy, we can express physical data in terms of derivative expansions of
fluid variables (fluid velocity $u(x)$, temperature $T(x)$ and chemical potential
 $\mu(x)$) around their equilibrium value.
 
The dynamics of the fluid is described by some conservation equations. 
For example, the conservation equations of the fluid stress-tensor or the fluid charge
current. These are known as constitutive equations.  
The stress tensor and charged current of fluid can be expressed in terms of fluid 
variables and their derivatives. At any derivative order, a generic form of 
the stress tensor and charged current can be written
demanding symmetry and thermodynamics of the underlying field theory. These generic expressions
are known as constitutive relations.  As it turns out, validity of 2nd law of thermodynamics
further constraints the form of these constitutive relations.

The author of \cite{Loganayagam:2011mu} assumed the following form for the
constitutive relations describing energy, charge and 
entropy transport in a fluid
\begin{equation}
\begin{split}
T^{\mu\nu} &\equiv \varepsilon u^\mu u^\nu + p P^{\mu\nu} + q^\mu_{anom}u^\nu + u^\mu q^\nu_{anom} + T^{\mu\nu}_{diss}\\
J^{\mu} &\equiv q u^\mu + J^{\mu}_{anom}+J^{\mu}_{diss} \\
J^\mu_S &\equiv s u^\mu + J^\mu_{S,anom}+J^\mu_{S,diss}\\
\end{split}
\end{equation}
where $u^\mu$ is the velocity of the fluid under consideration which obeys $u^\mu u_\mu =-1$ when
contracted using the space time metric $g_{\mu\nu}$. Further, $P^{\mu\nu}\equiv g^{\mu\nu}+u^\mu u^\nu$ , 
pressure of the fluid is $p$ and $\{\epsilon,q,s\}$ are the 
energy,charge and the entropy densities respectively. We have denoted by $\{q^\mu_{anom},J^{\mu}_{anom},
J^\mu_{S,anom}\}$ the anomalous heat/charge/entropy currents and by $\{T^{\mu\nu}_{diss},J^{\mu}_{diss},
J^\mu_{S,diss}\}$ the dissipative currents.

\subsubsection{Equation for adiabaticity}
A convenient way to describe adiabatic transport process is via a 
\textbf{covariant} anomalous Gibbs current $\prn{\mathcal{G}^{Cov}_{anom}}^\mu$. 

The adjective \textbf{covariant} refers to the fact that the Gibbs free energy
and the corresponding partition function are computed by turning on chemical 
potential for the \textbf{covariant} charge.  This is to be contrasted with 
the \textbf{consistent} partition function and
the corresponding \textbf{consistent} anomalous Gibbs current 
$\prn{\mathcal{G}^{Consistent}_{anom}}^\mu$.

Since this distinction is crucial let us elaborate this in the 
next few paragraphs - it is a fundamental result due to Noether that the continuous 
symmetries of a theory are closely linked to the conserved
currents in that theory. Hence when the path integral fails to have
a symmetry in the presence of background sources, there are 
two main consequences - first of all it directly
leads to a modification of the corresponding charge conservation and
a failure of Noether theorem. The second consequence is that various
correlators obtained by varying the path integral are not gauge-covariant
and a more general modifications of Ward identities occur.

A simple example is the expectation value of the current obtained by
varying the path integral  with respect to a gauge field (often termed
the \textbf{consistent} current ) as, 

\[ J^{\mu}_{Consistent}\equiv \frac{\partial S}{\partial {\cal A}_{\mu}} .\]
The consistent current is not covariant under gauge transformation. 

As has been explained in great detail in \cite{Bardeen:1984pm} thus
there exists another current in anomalous theories: the covariant current. 
The covariant current $J^{\mu}_{Cov}$ is a current shifted with respect 
to the consistent current by an amount $J_c^{\mu}$. The shift is such 
that its gauge transformation is anomalous and it exactly cancels the 
gauge non invariant part of the consistent current. Thus, the covariant
current is covariant under the gauge transformation, as suggested by its name.

The covariant Gibbs current describes the transport of Gibbs free energy when
a chemical potential is turned on for the covariant charge.
We will take a Hodge-dual of this covariant Gibbs current to get a $d-1$ form 
in d-space time dimensions. Let us denote this Hodge-dual by $\bar{\mathcal{G}}^{Cov}_{anom}$.
The anomalous parts of charge/entropy/energy currents can be derived from this 
Gibbs current via thermodynamics 
\begin{equation}
\begin{split}
\bar{J}^{Cov}_{anom} &= -\frac{\partial\bar{\mathcal{G}}_{anom}}{\partial\mu}\\
\bar{J}^{Cov}_{S,anom} &= -\frac{\partial\bar{\mathcal{G}}_{anom}}{\partial T}\\
\bar{q}^{Cov}_{anom} &= \bar{\mathcal{G}}_{anom} + T \bar{J}_{S,anom} + \mu \bar{J}_{anom}
\end{split}
\end{equation}

Then according to \cite{Loganayagam:2011mu} the condition for adiabaticity is
\begin{equation}\label{adiabiticity}
d\bar{q}^{Cov}_{anom} + \mathfrak{a} \wedge \bar{q}^{Cov}_{anom} -\mathcal{E}\wedge \bar{J}^{Cov}_{anom}
= T d\bar{J}^{Cov}_{S,anom} + \mu d\bar{J}^{Cov}_{anom} -\mu \bar{\mathfrak{A}}^{Cov}
\end{equation}
where $\mathfrak{a},\mathcal{E}$ are the acceleration 1-form and the rest-frame 
electric field 1-form respectively
defined via
\[ \mathfrak{a} \equiv (u.\nabla)u_\mu\ dx^\mu\ ,\quad \mathcal{E}\equiv u^\nu\mathcal{F}_{\mu\nu} dx^\mu  \] 
Further the rest frame magnetic field/vorticity 2-forms are defined by subtracting out
the electric part from the gauge field strength
and the acceleration part from the exterior derivative of velocity, viz.,
\[  \mathcal{B}\equiv \mathcal{F}-u\wedge\mathcal{E} \ ,\quad 2\omega \equiv du+u\wedge \mathfrak{a} \]

The symbol $\bar{\mathfrak{A}}^{Cov}$ is the d-form which is the 
Hodge dual of the rate at which the \textbf{covariant}
charge is created due to anomaly,i.e.,
\[ d\bar{J}^{Cov} = \bar{\mathfrak{A}}^{Cov} \]
where $\bar{J}^{Cov}$ is the entire covariant charge current including both the anomalous 
and the non-anomalous pieces. For simplicity we have restricted our attention to a single U(1)
global symmetry which becomes anomalous on a non-trivial background.

In terms of the Gibbs current , we can write the adiabiticity condition \eqref{adiabiticity} as,
\begin{equation}\label{eq:adiabG}
d\bar{\mathcal{G}}^{Cov}_{anom} + \mathfrak{a} \wedge \bar{\mathcal{G}}^{Cov}_{anom}+\mu \bar{\mathfrak{A}}^{Cov}
= \prn{dT+\mathfrak{a}T}\wedge \frac{\partial\bar{\mathcal{G}}^{Cov}_{anom}}{\partial T}
+ \prn{d\mu+\mathfrak{a}\mu-\mathcal{E}}\wedge \frac{\partial\bar{\mathcal{G}}^{Cov}_{anom}}{\partial \mu}
\end{equation}

\subsubsection{Construction of the polynomial $\mathfrak{F}^\omega_{anom}$}
The main insight of \cite{Loganayagam:2011mu}  is that in d-space time
dimensions the solutions  of this equation are most conveniently phrased 
in terms of a single homogeneous polynomial of degree $n+1$
in temperature $T$ and chemical potential $\mu$. 

Following the notation employed in \cite{Loganayagam:2012pz}
we will denote this polynomial as $\mathfrak{F}^\omega_{anom}[T,\mu]$. As was realised 
in \cite{Loganayagam:2012pz}, this polynomial is often closely related to the anomaly polynomial
of the system\footnote{We remind the reader that the anomalies of a theory living in $d=2n$ spacetime
dimensions is succinctly captured by a $2n+2$ form living in \emph{two dimensions higher}. This $2n+2$ form called the
anomaly polynomial (since it is a polynomial in external/background field strengths $\mathcal{F}$ and $\mathfrak{R}$)
is related to the variation of the effective action $\delta W$ via \emph{the descent relations} 
\[ \mathcal{P}_{anom}=d\Gamma_{CS}\ ,\qquad \delta \Gamma_{CS}= d \delta W \]
We will refer the reader to various textbooks\cite{Weinberg:1996kr,
Bertlmann:1996xk,Bastianelli:2006rx} and lecture notes \cite{Harvey:2005it,Bilal:2008qx} for a more detailed
exposition.} . More precisely, for a variety  of systems we have a remarkable relation
between $\mathfrak{F}_{anom}^\omega[T,\mu]$  and the anomaly polynomial $\mathcal{P}_{anom} \brk{ \mathcal{F}, \mathfrak{R}}$ 
\begin{equation}\label{eq:anomFP}
\begin{split}
\mathfrak{F}_{anom}^\omega[T,\mu] = \mathcal{P}_{anom} \brk{ \mathcal{F}  \mapsto \mu, p_1(\mathfrak{R}) \mapsto - T^2 , p_{k>1}(\mathfrak{R}) \mapsto 0 }
\end{split}
\end{equation}
Let us be more specific : on a $(2n-1)+1$ dimensional space time  consider a theory with 
\begin{equation}\label{eq:FOmegaC}
\begin{split}
\mathfrak{F}^\omega_{anom}[T,\mu] &= \mathcal{C}_{anom}\mu^{n+1}+\sum_{m=0}^{n}C_m T^{m+1}\mu^{n-m}\\
\end{split}
\end{equation}
Assuming that the theory obeys the 
replacement rule \eqref{eq:anomFP} such a $\mathfrak{F}^\omega_{anom}[T,\mu]$ can be obtained
from an anomaly polynomial\footnote{Since all
relativistic theories only have integer powers of Pontryagin forms the constants $C_{m}$ should
vanish whenever $m$ is even. As we shall see later that another way to arrive at the same conclusion 
is to impose CPT invariance.}
\begin{equation}
\begin{split}
\mathcal{P}_{anom} &= \mathcal{C}_{anom}\mathcal{F}^{n+1}+\sum_{m=0}^{n}C_m \brk{- p_1(\mathfrak{R})}^{\frac{m+1}{2}}\mathcal{F}^{n-m}+\ldots\\
\end{split}
\end{equation}
where we have presented the terms which do not involve the higher Pontryagin forms. 
Restricting our attention only to the $U(1)^{n+1}$ anomaly (and ignoring the mixed/pure
gravitational anomalies ) we can write 
\begin{equation}\label{eq:dJ}
 \begin{split}
  d\bar{J}_{Consistent} &=\mathcal{C}_{anom}\mathcal{F}^n\\
 d\bar{J}_{Cov} &=(n+1)\mathcal{C}_{anom} \mathcal{F}^n \\
 \end{split}
\end{equation}
and their difference is given by
\begin{equation}\label{eq:shift}
 \begin{split} 
  \bar{J}_{Cov} = \bar{J}_{Consistent}+n \mathcal{C}_{anom}\hat{\mathcal{A}}\wedge \mathcal{F}^{n-1}
 \end{split}
\end{equation}

The solution of \eqref{eq:adiabG} corresponding to the homogeneous polynomial \eqref{eq:FOmegaC}
is given by
\begin{equation}\label{eq:GCovBOmega}
\begin{split}
\bar{\mathcal{G}}^{Cov}_{anom} 
&= C_0 T \hat{\mathcal{A}}\wedge\mathcal{F}^{n-1}+ \sum_{m=1}^{n}\left[\mathcal{C}_{anom}\binom{n+1}{m+1}\mu^{m+1}\right.\\
&\qquad  \left. + \sum_{k=0}^{m}C_k \binom{n-k}{m-k}  T^{k+1}\mu^{m-k}\right] (2\omega)^{m-1} \mathcal{B}^{n-m}\wedge u \\
\end{split}
\end{equation}
Here $\hat{\mathcal{A}}$ is the $U(1)$ gauge-potential 1-form in some gauge with
$\mathcal{F}\equiv d\hat{\mathcal{A}}$ being its field-strength 2-form. Further,
$\mathcal{B},\omega$ are the rest frame magnetic field/vorticity 2-forms 
and  $T ,\mu $ are the local temperature and  chemical potential respectively. 
They obey
\begin{equation}
\begin{split}
(d\mathcal{B})\wedge u  = -(2\omega)\wedge\mathcal{E}\wedge u \ , \quad d(2\omega)\wedge u = (2\omega)\wedge\mathfrak{a}\wedge u
\end{split}
\end{equation}
Using these equations  it is a straightforward exercise to check that
\eqref{eq:GCovBOmega} furnishes a solution to \eqref{eq:adiabG}.

We will make a few remarks before we proceed to derive charge/entropy/energy currents
from this Gibbs current. Note that if one insists that the Gibbs current be gauge-invariant
then we are forced to  put $C_0=0$ - in the solution presented in 
\cite{Loganayagam:2011mu}  this condition was implicitly assumed 
and the $C_0$ term was absent. The authors of \cite{Banerjee:2012iz} later
relaxed this assumption  insisting gauge-invariance only for 
the covariant charge/energy currents. Since we would be interested in 
comparison with the results derived in \cite{Banerjee:2012iz} it 
is useful to retain the $C_0$ term.

Now we use thermodynamics to obtain the charge current as
\begin{equation}
\begin{split}
&\bar{J}^{Cov}_{anom} \\ 
&=- \sum_{m=1}^{n} \left[ (m+1)\mathcal{C}_{anom}\binom{n+1}{m+1} \mu^m 
\right.\\
&\qquad\left. +\sum_{k=0}^{m}(m-k)C_k \binom{n-k}{m-k}  T^{k+1}\mu^{m-k-1}\right] (2\omega)^{m-1} \mathcal{B}^{n-m}\wedge u \\
\end{split}
\end{equation}
and the entropy current is given by
\begin{equation}
\begin{split}
\bar{J}^{Cov}_{S,anom}
&=- C_0\hat{\mathcal{A}}\wedge\mathcal{F}^{n-1}\\
&\qquad - \sum_{m=1}^{n}\sum_{k=0}^{m}(k+1) C_k \binom{n-k}{m-k}  T^{k}\mu^{m-k} (2\omega)^{m-1} \mathcal{B}^{n-m}\wedge u\\
\end{split}
\end{equation}
The energy current is given by 
\begin{equation}
\begin{split}
&\bar{q}^{Cov}_{anom}\\ 
&=- \sum_{m=1}^{n}m\left[\mathcal{C}_{anom}\binom{n+1}{m+1} \mu^{m+1}
\right.\\
&\qquad \left.+\sum_{k=1}^{m}C_k \binom{n-k}{m-k}  T^{k+1}\mu^{m-k}\right] (2\omega)^{m-1} \mathcal{B}^{n-m}\wedge u \\
\end{split}
\end{equation}

These currents satisfy an interesting Reciprocity type relationship
noticed in \cite{Loganayagam:2011mu}
\begin{equation}\label{eq:reciprocity}
\frac{\delta \bar{q}^{Cov}_{anom}}{\delta \mathcal{B}} = \frac{\delta \bar{J}^{Cov}_{anom}}{\delta (2\omega)}
\end{equation}

While this is a solution in a generic frame one can specialise to the Landau frame (where the
velocity is defined via the energy current) by a frame transformation 
\begin{equation}
\begin{split}
u^\mu &\mapsto u^\mu - \frac{q^\mu_{anom}}{\epsilon + p}, \\
J^\mu_{anom} &\mapsto J^\mu_{anom} - q \frac{q^\mu_{anom}}{\epsilon + p} ,\\
J^\mu_{S,anom}& \mapsto J^\mu_{S,anom} - s \frac{q^\mu_{anom}}{\epsilon + p},\\
q^\mu_{anom} &\mapsto 0\\
\end{split}
\end{equation}
to get
\begin{equation}
\begin{split}
\bar{J}^{Cov,Landau}_{anom} &= \sum_{m=1}^{n}\xi_m(2\omega)^{m-1} \mathcal{B}^{n-m}\wedge u\\ 
\bar{J}^{Cov,Landau}_{S,anom} &= \sum_{m=1}^{n}\xi^{(s)}_m(2\omega)^{m-1} \mathcal{B}^{n-m}\wedge u+\zeta\ \hat{\mathcal{A}}\wedge\mathcal{F}^{n-1}\\ 
\end{split}
\end{equation}
where
\begin{equation}\label{eq:xiLoga}
\begin{split}
\xi_m &\equiv \brk{m \frac{q\mu}{\epsilon + p}-(m+1)}\mathcal{C}_{anom}\binom{n+1}{m+1} \mu^m\\
&\qquad+\sum_{k=0}^{m}\brk{m \frac{q\mu}{\epsilon + p}-(m-k)}C_k \binom{n-k}{m-k}  T^{k+1}\mu^{m-k-1} \\
\xi^{(s)}_m &\equiv \brk{m \frac{sT}{\epsilon + p}}\mathcal{C}_{anom}\binom{n+1}{m+1} T^{-1}\mu^{m+1}\\
&\qquad+\sum_{k=0}^{m}\brk{m \frac{sT}{\epsilon + p}-(k+1)}C_k \binom{n-k}{m-k}  T^k\mu^{m-k} \\
\zeta &= - C_0
\end{split}
\end{equation}
Often in the literature the entropy current is quoted in the form 
\begin{equation}
\begin{split}
\bar{J}^{Cov,Landau}_{S,anom} &= -\frac{\mu}{T} \bar{J}^{Cov,Landau}_{anom} + \sum_{m=1}^{n}\chi_m(2\omega)^{m-1} \mathcal{B}^{n-m}\wedge u+\zeta\ \hat{\mathcal{A}}\wedge\mathcal{F}^{n-1}\\ 
\end{split}
\end{equation}
where
\begin{equation}\label{eq:Chi_mPrediction}
\begin{split}
\zeta &= -C_0 \\ 
\chi_m &\equiv \xi^{(s)}_m +\frac{\mu}{T} \xi_m\\
&= - \mathcal{C}_{anom}\binom{n+1}{m+1} T^{-1}\mu^{m+1}-\sum_{k=0}^{m}C_k \binom{n-k}{m-k}  T^k\mu^{m-k} \\
\end{split}
\end{equation}
where we have used the thermodynamic relation $sT+q\mu=\epsilon + p$. By looking at \eqref{eq:GCovBOmega}
we recognise these to be the coefficients occurring in the anomalous Gibbs current :
\begin{equation}\label{eq:GibbsChi}
\begin{split}
\bar{\mathcal{G}}^{Cov}_{anom} &= -T\brk{\sum_{m=1}^{n}\chi_m(2\omega)^{m-1} \mathcal{B}^{n-m}\wedge u+\zeta\ \hat{\mathcal{A}}\wedge\mathcal{F}^{n-1}}\\ 
\end{split}
\end{equation}
In fact this is to be expected from basic thermodynamic considerations : the above equation is
a direct consequence of the relation $G=-T(S+\frac{\mu}{T}Q-\frac{U}{T})$ and the fact that 
energy current receives no anomalous contributions in the Landau frame.

This ends our review of the main results of \cite{Loganayagam:2011mu} adopted to 
our purposes. Our aim in the rest of the paper would be to derive all these results 
purely from a partition function analysis. 

\subsection{Equilibrium Partition Function}\label{subsec:PartitionReview}

In this subsection we review (and extension) an alternative approach to constrain the
constitutive relations, namely by demanding the existence  of an equilibrium
partition function (or free energy) for the fluid  as described in \cite{Banerjee:2012iz,Jain:2012rh}
\footnote{For similar discussions, see for example
\cite{Jensen:2012jh,Jensen:2012jy}.}.

Let us keep the fluid in a special background such that the background metric has a time
like killing vector and the background gauge field is time independent. 
Any such metric can be put into the following Kaluza-Klein form
\begin{equation}
\begin{split}
ds^2 &= -e^{2\sigma}(dt+a_idx^i)^2 + g_{ij}dx^idx^j, \\
\hat{\cal A} &= {\cal A}_0dt + {\cal A}_idx^i
\end{split}
\label{KKform}
\end{equation}
here $i,j~\epsilon ~~(1,2 \ldots 2n-1)$ are the spatial indices. We will often use the 
notation $\gamma\equiv e^{-\sigma}$ for brevity. This background has a time-like
killing vector $\partial_t$ and let $u_k^\mu=(e^{-\sigma},0,0,\ldots)$ be the 
unit normalized vector in the killing direction so that
\[  u_k^\mu\partial_\mu = \gamma \partial_t \quad\text{and}\quad  u_k= -\gamma^{-1}(dt+a) \]

In the corresponding Euclidean field theory description of equilibrium, the imaginary
time direction would be compactified into a thermal circle with the 
size of circle being the inverse temperature of the underlying
field theory. In the 2n-1 dimensional compactified geometry, the original 2n background field
breaks as follow
\begin{itemize}
\item metric($g_{\mu\nu}$) : scalar($\sigma$), KK gauge field($a_i$), lower dimensional metric($g_{ij}$).
\item gauge field($\hat{\cal A}_\mu$) : scalar(${\cal A}_0$), gauge field(${\cal A}_i$)
\end{itemize}

Under this KK type reduction the 2n dimensional diffeomorphisms breaks up into
2n-1 dimensional diffeomorphisms and KK gauge transformations. The components
of 2n dimensional tensors which are KK-gauge invariant in 2n-1 dimensions are those with lower
time(killing direction) and upper space indices. Given a 1-form $J$ 
we will split it in terms of KK-invariant components as 
\[ J=J_0 (dt+a_i dx^i)+ g_{ij} J^i dx^j \]
Other KK non-invariant components of $J$ are given by 
\begin{equation}
\begin{split}
J^0 &= -\brk{\gamma^2 J_0+a_i J^i}\\ 
J_i &= g_{ij}J^j +a_i J_0
\end{split}
\end{equation}

To take care of KK gauge invariance we will identify the lower dimensional U(1) 
gauge field (denoted by non script letters) as follows
\begin{equation}
\begin{split}
A_0 &= {\cal A}_0+\mu_0 , ~~A^i = {\cal A}^i \\
\Rightarrow A_i &= {\cal A}_i - {\cal A}_0 a_i {\rm ~~~and}\\
F_{ij} &= \partial_i A_j - \partial_j A_i =  \mathcal{F}_{ij}
            - A_0 f_{ij} -(\partial_i A_0~a_j - \partial_j A_0~a_i).
\end{split}
\label{KKinv}
\end{equation}
where $f_{ij}\equiv \partial_i a_j - \partial_j a_i$ and $\mu_0$ is a convenient 
constant shift in ${\cal A}_0$ which we will define shortly. We can hence write
\[\hat{\mathcal{A}} = \mathcal{A}_0 dt + \mathcal{A} = A_0 (dt+a_i dx^i) + A_i dx^i-\mu_0 dt \]
We are now working in a general gauge - often it is useful to
work in a specific class of gauges : one class of gauges we will work on is obtained from this generic gauge by 
performing a gauge transformation to remove the $\mu_0 dt$ piece. We will call these class
of gauges as the `zero $\mu_0$' gauges. In these gauges the new gauge field is given in terms of the
old gauge field via 
\[ \hat{\mathcal{A}}_{\mu_0=0} \equiv  \hat{\mathcal{A}}+\mu_0 dt \]
We will quote all our consistent currents in this gauge. The field strength 2-form can then be written as
\[ \mathcal{F}\equiv d\hat{\mathcal{A}} = dA+ A_0 da + dA_0\wedge(dt+a) \]

We will now focus our attention on the \textbf{consistent} equilibrium partition function
which is the Euclidean path-integral computed on space adjoined with a thermal circle of
length $1/T_0$. We will further turn on a chemical potential $\mu$ - since there are various
different notions of charge in anomalous theories placed in gauge backgrounds we need to 
carefully define which of these notions we use to define the partition function\footnote{
See, for example, section\S 3 of  \cite{Landsteiner:2011tf} for a discussion of some of the 
subtleties.}. While in the previous subsection we used the chemical potential for a 
\textbf{covariant} charge and the corresponding \textbf{covariant} Gibbs free-energy
following \cite{Loganayagam:2011mu} , in this subsection we will follow \cite{Banerjee:2012iz}
in using a chemical potential for the consistent charge to define the partition 
function. This distinction has to be kept in mind while making a comparison between the
two formalisms as we will elaborate later in section\S\ref{sec:IntByParts}.

The consistent partition function $Z_{Consistent}$ that we write down will be 
the most general one consistent with 2n-1 dimensional diffeomorphisms, 
KK gauge invariance and the U(1) gauge invariance up to anomaly. 
It is a scalar $S$ constructed out of various background quantities and their derivatives.
The most generic form of the partition function is
\begin{equation}\label{parform}
W=\ln Z_{Consistent}= \int d^{2n-1}x \sqrt{g_{2n-1}} S(\sigma, A_0,a_i,A_i,g_{ij}) .
\end{equation}
Given this partition function, we compute various components of the stress tensor and 
charged current from it. The KK gauge invariant components of the stress tensor 
$T_{\mu\nu}$ and charge current $J_{\mu}$ can then be 
obtained from the partition function as follows \cite{Banerjee:2012iz},
\begin{equation}\label{parstcu}
\begin{split}
T_{00} &= -\frac{T_0 e^{2 \sigma}}{\sqrt{-g_{2n}} }\frac{\delta W}{\delta \sigma},~~
J_0^{Consistent} = -\frac{e^{2\sigma} T_0}{\sqrt{-g_{2n}}}\frac{\delta W}{\delta A_0}, \\
T_0^i &= \frac{T_0}{\sqrt{-g_{2n}} }\bigg(\frac{\delta W}{\delta a_i}
       - A_0 \frac{\delta W}{\delta A_i}\bigg),~~
J^i_{Consistent} = \frac{T_0}{\sqrt{-g_{2n}}}\frac{\delta W}{\delta A_i}, \\~~
T^{ij} &= -\frac{2 T_0}{\sqrt {-g_{2n}}} g^{il}g^{jm}\frac{\delta W}
           {\delta g^{lm}}. \\
\end{split}
\end{equation}
here $\{ \sigma, a_i, g_{ij}, A_0, A_i \}$ are chosen independent sources, so the partial
derivative w.r.t any of them in the above equations means that others are kept constant. 
We will sometimes use the above equation written in terms of differential forms - we will
refer the reader to appendix~\ref{app:variationForms}  for the differential-form version 
of the above equations.

Next 
we parameterize the most generic equilibrium solution and constitutive relations for the fluid as,
\begin{eqnarray}\label{flustcu}
&& u(x)= u_0(x)+u_1(x) , \quad T(x)= T_0(x)+T_1(x) , \quad \mu(x)= \mu_0(x)+\mu_1(x) , \nonumber \\
&& T_{\mu\nu}=(\epsilon + p)u_{\mu}u_{\nu} + p g_{\mu\nu}+\pi_{\mu\nu} , \quad J^{\mu}= 
q u^{\mu}+ j^{\mu}_{diss},
\end{eqnarray}
where, ${u_1,T_1,\mu_1,\pi_{\mu\nu},j^{\mu}_{diss}}$ are various derivatives of the background 
quantities. Note that we will work in Landau frame throughout.

These corrections are found by comparing the fluid stress tensor 
$T_{\mu\nu}$ and current $J_{\mu}$ in Eqn.\eqref{flustcu} with $T_{\mu\nu}$ and $J_{\mu}$ 
in Eqn.\eqref{parstcu} as obtained from the partition function.  This exercise
then constrains various non-dissipative coefficients that appear in the 
constitutive relations in Eqn.\eqref{flustcu}.

This then ends our short review of the formalism developed in \cite{Banerjee:2012iz}.
In the next section we will apply this formalism to a theory with $U(1)^{n+1}$ 
anomaly in $d=2n$ space time dimensions.

\section{Anomalous partition function in arbitrary dimensions}\label{sec:2ndimu1}
Let us consider then a fluid  in a $2n$ dimensional space time. The fluid is
charged under a single $ U(1)$ abelian gauge field ${\cal A}_{\mu}$. We will
generalise to multiple abelian gauge fields  later in section \S\ref{sec:2ndimmul}
and leave the non-abelian case for future study. We will continue to use the
notation in the subsection \S\S\ref{subsec:LogaReview}.

The consistent/covariant anomaly are then given by Eqn.\eqref{eq:dJ} which can 
be written in components as
\begin{equation}
\begin{split}
\nabla_{\mu}J^{\mu}_{Consistent} &= \mathcal{C}_{anom} 
\varepsilon^{\mu_1 \nu_1 \ldots\mu_n\nu_n}\partial_{\mu_1}\hat{\mathcal{A}}_{\nu_1}\ldots\partial_{\mu_n}\hat{\mathcal{A}}_{\nu_n}\\
&= \frac{\mathcal{C}_{anom}}{2^n} \varepsilon^{\mu_1 \nu_1 \ldots\mu_n\nu_n} \mathcal{F}_{\mu_1 \nu_1} \ldots \mathcal{F}_{\mu_n \nu_n}.\\
\nabla_{\mu}J^{\mu}_{Cov} &= (n+1) \mathcal{C}_{anom} 
\varepsilon^{\mu_1 \nu_1 \ldots\mu_n\nu_n}\partial_{\mu_1}\hat{\mathcal{A}}_{\nu_1}\ldots\partial_{\mu_n}\hat{\mathcal{A}}_{\nu_n}\\
&= (n+1) \frac{\mathcal{C}_{anom}}{2^n} \varepsilon^{\mu_1 \nu_1 \ldots\mu_n\nu_n} \mathcal{F}_{\mu_1 \nu_1} \ldots \mathcal{F}_{\mu_n \nu_n}.
\end{split}
\end{equation}
and Eqn.\eqref{eq:shift} becomes
\begin{equation}\label{covcur}
J_{Cov}^{\mu} = J^{\mu}_{Consistent} + J^{\mu}_{(c)}.
\end{equation}
where
\begin{equation}\label{curcor}
\begin{split}
J^{\lambda}_{(c)} &= n\mathcal{C}_{anom} \varepsilon^{\lambda \alpha\mu_1 \nu_1 \ldots\mu_{n-1}\nu_{n-1}}
\hat{\mathcal{A}}_{\alpha} \partial_{\mu_1}\hat{\mathcal{A}}_{\nu_1}\ldots\partial_{\mu_{n-1}}\hat{\mathcal{A}}_{\nu_{n-1}}\\
&=n\frac{\mathcal{C}_{anom}}{2^{n-1}} \varepsilon^{\lambda \alpha\mu_1 \nu_1 \ldots\mu_{n-1}\nu_{n-1}}
\hat{\mathcal{A}}_{\alpha}  \mathcal{F}_{\mu_1 \nu_1} \ldots \mathcal{F}_{\mu_{n-1} \nu_{n-1}}.
\end{split}
\end{equation}

The energy-momentum equation becomes
\begin{equation}
\nabla_{\mu}T^{\mu}_{\nu}= F_{\nu \mu}J^{\mu}_{Cov} ,
\end{equation}
where $J^{\mu}_{Cov}$ is the  covariant current. This has been
explicitly shown in \cite{Banerjee:2012iz} \footnote{One required identity
is, 
\[ \hat{\mathcal{A}}_{\alpha}
 \varepsilon^{\mu_1 \nu_1 \ldots\mu_n\nu_n} \mathcal{F}_{\mu_1 \nu_1} \ldots \mathcal{F}_{\mu_n \nu_n}
= 2n\    \hat{\mathcal{A}}_{\mu_1}
 \varepsilon^{\mu_1 \nu_1 \mu_2\nu_2 \ldots\mu_n\nu_n} \mathcal{F}_{\alpha \nu_1} \mathcal{F}_{\mu_2 \nu_2}\ldots \mathcal{F}_{\mu_n \nu_n}
\]
for arbitrary $2n-$dimensions}. 

\subsection{Constraining the partition function}
We want to write the equilibrium free energy functional for the fluid. For this purpose,
let us keep the in the following $2n$-dimensional time independent background,
\begin{eqnarray}\label{backgr}
ds^2= - e^{2 \sigma}(dt+ a_i dx^i)^2+ g_{ij}dx^idx^j, \quad {\cal A}= (A_0, {\cal A}_i).
\end{eqnarray}

Now, we write the $(2n-1)$ dimensional equilibrium free energy that reproduces the same
anomaly as given in \eqref{anomeq}. The most generic form for the anomalous part of 
the partition function is ,
\begin{equation}\label{action}
\begin{split}
W_{anom}&=\frac{1}{T_0}\int d^{2n-1}x  \sqrt{g_{2n-1}}\bigg\{ \sum_{m=1}^{n}\alpha_{m-1}(A_0,T_0)\
 \brk{\epsilon A (da)^{m-1}(dA)^{n-m}}  \bigg.\\
&\qquad \bigg.\qquad  + \alpha_{n}(T_0)\  \brk{\epsilon a (da)^{n-1}} \bigg\}.
\end{split}
\end{equation}
where, $\epsilon^{ijk\ldots }$ is the $(2n-1)$ dimensional tensor density defined via
\[ \epsilon^{i_1i_2\ldots i_{d-1}} = e^{-\sigma}\varepsilon^{0i_1i_2\ldots i_{d-1}} \]
The indices $(i,j)$ run over $(2n-1)$ values. We have used the following notation
for  the sake of brevity
\begin{equation}\label{epsDef}
\begin{split}
&\brk{\epsilon A (da)^{m-1}(dA)^{n-m}} \\
&\quad \equiv \epsilon^{i j_1k_1 \ldots j_{m-1} k_{m-1} p_1 q_1\ldots p_{n-m}q_{n-m}}
A_i \partial_{j_1} a_{k_1}\ldots \partial_{j_{m-1}} a_{k_{m-1}}\partial_{p_1} A_{q_1}\ldots \partial_{p_{n-m}} A_{q_{n-m}}\\
&\brk{\epsilon (da)^{m-1}(dA)^{n-m}}^i \\
&\quad \equiv \epsilon^{i j_1k_1 \ldots j_{m-1} k_{m-1} p_1 q_1\ldots p_{n-m}q_{n-m}}
\partial_{j_1} a_{k_1}\ldots \partial_{j_{m-1}} a_{k_{m-1}}\partial_{p_1} A_{q_1}\ldots \partial_{p_{n-m}} A_{q_{n-m}}\\
\end{split}
\end{equation}

The invariance under diffeomorphism implies that $\alpha_{n}$ is a
constant in space .For $m<n$ however $\alpha_m$ can have $A_0$ dependence, as the gauge 
symmetry is anomalous, but they are  independent of $\sigma$, due to diffiomorphism invariance.

The consistent current computed from this partition function is,
\begin{equation}
\begin{split}
\prn{J_{anom}}_0^{Consistent} &=- e^{\sigma}\sum_{m=1}^{n} \frac{\partial\alpha_{m-1}}{\partial A_0}\brk{\epsilon A (da)^{m-1}
(dA)^{n-m}}  \\
\prn{J_{anom}}^i_{Consistent} &=e^{-\sigma} \bigg\{\sum_{m=1}^{n}    (n-m+1) \alpha_{m-1}
\brk{\epsilon(da)^{m-1}(dA)^{n-m}}^i  \bigg.  \\
&\bigg. - \sum_{m=1}^{n-1} (n-m)  \frac{\partial\alpha_{m-1}}{\partial A_0} \brk{\epsilon A dA_0(da)^{m-1}(dA)^{n-m-1}}^i \bigg\}
\end{split}
\end{equation}

Next, we compute the covariant currents, following \eqref{covcur}. The correction piece for the 0-component of the current is,
\begin{equation}
(J_{(c)})_0=-n\mathcal{C}_{anom} e^{\sigma}
\sum_{m=1}^{n}  A_0^{m}\binom{n-1}{m-1} \brk{\epsilon A (da)^{m-1}
(dA)^{n-m}}
\end{equation}
where, we have used the following identification for $2n$ dimensional gauge field
${\cal A}_{\mu}$ and $(2n-1)$ dimensional gauge fields $A_i, a_i$ and scalar $A_0$,
\begin{equation}
 \begin{split}
  \mathcal A_i &= A_i + a_i A_0  \\
\mathcal A_0 &=  A_0.
 \end{split}
\end{equation}
where we are working in a`zero $\mu_0$' gauge. 

Thus, the 0-component of the covariant current is,
\begin{equation}
\prn{J_{anom}}_0^{Cov}= -e^{\sigma} \epsilon^{ijkl\ldots}\sum_{m=1}^{n} \brk{  \frac{\partial\alpha_{m-1}}{\partial A_0} +
n\binom{n-1}{m-1} A_0^{m-1} \mathcal{C}_{anom} }
\brk{\epsilon A (da)^{m-1}(dA)^{n-m}}.
\end{equation}
Every term in the above sum is  gauge non-invariant. So the covariance of the covariant current
demands that  we chose the arbitrary functions $\alpha_m$ appearing in the partition function \eqref{action}
such that the current vanishes. Thus, we get,
\begin{equation}
\frac{\partial\alpha_{m-1}}{\partial A_0} +
n\binom{n-1}{m-1} A_0^{m-1} \mathcal{C}_{anom}  =0 .
\end{equation}
The solution for the above equation is,
\begin{eqnarray}\label{Csol}
&&\alpha_m = - \mathcal{C}_{anom}\binom{n}{m+1} A_0^{m+1} + \tilde{C}_m T_0^{m+1}  ,
\quad m=0, \ldots, n-1 \nn\\
&&\alpha_n =  \tilde{C}_n T_0^{n+1}
\end{eqnarray}
Here, $\tilde C_m$ are constants that can appear in the partition function.

Thus, at this point, a total of $n+1$ coefficients can appear in the partition function. 
A further study of CPT invariance of the partition function will reduce this number. We will present that analysis later
in details and here we just state the result. CPT forces all $\tilde{C}_{2k} =0$.
For even $n$, the number of constants are $\frac{n}{2}$ where as for odd $n$, 
the number is $(\frac{n+1}{2})$.

\subsection{Currents from the partition function}
With these functions the $i-$component of the covariant current is,
\begin{equation}\label{cufp}
\begin{split}
\prn{J_{anom}}_{Cov}^i  
&=  e^{-\sigma} \sum_{m=1}^{n} \bigg[A_0 \frac{\partial\alpha_{m-1}}{\partial A_0} +(n-m+1)\alpha_{m-1} \bigg]\brk{\epsilon (da)^{m-1} (dA)^{n-m}}^i \\
&=  e^{-\sigma} \sum_{m=1}^{n} \bigg[-(n+1) \mathcal{C}_{anom}
\binom{n}{m} T_0 A_0^{m}\bigg.\\
&\qquad \bigg. \qquad +(n-m+1)T_{0}^{m}\tilde C_{m-1} \bigg]\brk{\epsilon (da)^{m-1} (dA)^{n-m}}^i ,
\end{split}
\end{equation}
As expected, this current is $U(1)$ gauge invariant. The different components of stress-tensor
computed from the partition function are,
\begin{equation}\label{stfp}
\begin{split}
T^{anom}_{00}&=0, \qquad T_{anom}^{ij}=0 \\
\prn{T^i_0}_{anom} &=  e^{-\sigma} \sum_{m=1}^{n}\prn{ m \alpha_{m} - (n-m+1) A_0
\alpha_{m-1}} \brk{\epsilon (da)^{m-1} (dA)^{n-m}}^i\\
&=  e^{-\sigma} \sum_{m=1}^{n}\left[m \tilde{C}_m T_0^{m+1}-(n+1-m) \tilde{C}_{m-1} T_0^m A_0\right.\\
&\qquad\left.\qquad+\binom{n+1}{m+1}\mathcal{C}_{anom}A_0^{m+1}\right]\brk{\epsilon (da)^{m-1} (dA)^{n-m}}^i\\
\end{split}
\end{equation}

\subsection{Comparison with Hydrodynamics}
Next, we find the equilibrium solution for the fluid variables. As usual, we keep the fluid in
the time independent background \eqref{backgr}. The equilibrium solutions for perfect charged
fluid (with out any dissipation) are,
\begin{equation}
u^{\mu}\partial_\mu= e^{-\sigma}\partial_t , \quad T=T_0 e^{-\sigma}, \quad \mu= A_0 e^{-\sigma} .
\end{equation}

The most generic constituitive relations for the fluid can be written as,
\begin{eqnarray}\label{conscor}
T_{\mu\nu}&=& (\epsilon+p) u_{\mu}u_{\nu} +p g_{\mu\nu} + \eta \sigma_{\mu\nu}+ \zeta \Theta
{\cal P}_{\mu\nu} \nonumber \\
J^{\mu}_{Cov}&=& q u^{\mu}+ J^{\mu}_{even} +J^{\mu}_{odd} ,\nonumber \\
J^{\mu}_{even}&=&\sigma (E^{\mu}- T {\cal P}^{\mu\alpha} \partial_{\alpha}\nu )+
\alpha_1 E^{\mu} + \alpha_2 T {\cal P}^{\mu\alpha} \partial_{\alpha}\nu  +
\mbox{higher derivative terms}\nonumber \\
J^{\mu}_{odd}&=&  \sum_{m=1}^{n} \xi_m \varepsilon^{\mu \nu\ \gamma_1\delta_1\ldots\gamma_{m-1}\delta_{m-1}\ \alpha_1\beta_1 \ldots\alpha_{n-m}\beta_{n-m}}
u_{\nu}(\partial_{\gamma}u_{\delta})^{m-1}(\partial_{\alpha}{\cal A}_{\beta})^{n-m}+\ldots.
\end{eqnarray}
Here, $J^{\mu}_{even}$ is parity even part of the  charge current and
$J^{\mu}_{odd}$ is parity odd  charge current. $\varepsilon^{\mu \nu \alpha
\beta \gamma \delta \ldots}$ is a $2n$ dimensional tensor density whose $(n-m)$ indices
are contracted with $\partial_{\alpha}{\cal A}_{\beta}$ and $(m-1)$ indices are contracted with 
$\partial_{\gamma}u_{\delta}$.

We notice that the higher derivative part of the current gets contribution from both parity even and
odd vectors. Parity even vectors can be at any derivative order but parity odd vectors
always appear at $(n-1)$ derivative order. Thus, for a generic value of $n$ (other than $n=2$) ,
the parity even and odd parts corrections to the  current will always appear at
different derivative orders. From now on, we will only concentrate on the parity odd sector.
It is also straight forward to check that $J_{0}^{odd}=0$.

Next, we look for the equilibrium solution for this fluid. Since, there exist no
gauge invariant parity odd scalar, the temperature and chemical potential do not get any correction.
Also, in $2n$ dimensional theory, the parity odd vectors that we can write are always $(n-1)$
derivative terms. No other parity odd vector at any lower derivative order exists. Since the
fluid velocity is always normalized to unity, we have,
\begin{equation}
\delta T=0, \quad \delta \mu=0, \quad \delta u_0 = - a_i \delta u^i.
\end{equation}
where, the most generic correction to the fluid velocity is,
\begin{equation}
\delta u^i = \sum_{m=1}^{n}  U_m(\sigma, A_0) \brk{\epsilon(da)^{m-1} (dA)^{n-m}}^i .
\end{equation}

Here, $U_m(\sigma, A_0)$ are arbitrary coefficients and factors of $e^{\sigma}$ is introduced for later convenience.
 Similarly, we can parameterize the $i-$component
of the parity-odd  current as,
\begin{equation}\label{jdiss}
J^{i}_{odd} =  \sum_{m=1}^{n} J_m(\sigma, A_0) \brk{\epsilon(da)^{m-1} (dA)^{n-m}}^i .
\end{equation}
The coefficients $J_m(\sigma, A_0)$ are related to the transport coefficients $\xi_m$ via 
\begin{equation}\label{eq:jxi}
J_m= \sum_{k=1}^m \binom{n-k}{m-k} \xi_k \prn{-e^\sigma}^{k-1} A_0^{m-k}  .
\end{equation}

With all these data, we can finally compute the corrections to the stress tensor and charged currents 
and they take the following form,
\begin{eqnarray}\label{stc}
\delta T_{00}&=&0, \quad \delta T^{ij}=0, \quad \delta \tilde J_{0}=0 \nonumber \\
\delta T_{0}^i&=& -e^{\sigma} (\epsilon+p) \epsilon^{ijk\ldots} \sum_{m=1}^{n}
U_m(\sigma, A_0) ( da)^{m-1} (dA)^{n-m}  \nonumber \\
\delta J^{i}_{Cov} &=& \epsilon^{ijk\ldots}  \sum_{m=1}^{n}(J_m(\sigma, A_0) +
q U_m(\sigma, A_0))( da)^{m-1} (dA)^{n-m}
\end{eqnarray}
Comparing the expressions for various components of stress tensor and covariant current of
the fluid obtained from equilibrium partition function \eqref{stfp}, \eqref{cufp} and fluid
constitutive relations \eqref{stc}, we get,
\begin{eqnarray}\label{velocur}
U_{m} &=& - \frac{e^{-2\sigma}}{\epsilon+p}\left[m \alpha_m-(n-m+1)A_0\alpha_{m-1}\right]\nn\\
&=&-\frac{e^{-2\sigma}}{\epsilon+p}\left[m \tilde{C}_m T_0^{m+1}-(n+1-m) \tilde{C}_{m-1}  A_0T_0^m\right.\nn\\
&&\qquad\left.\qquad+\binom{n+1}{m+1}\mathcal{C}_{anom}A_0^{m+1}\right]
\end{eqnarray}
 Similarly, we can evaluate
$J_m(\sigma, A_0) $ as follows,
\begin{equation}\label{Jm}
\begin{split}
J_{m}&=e^{-\sigma}\brk{-(m+1)\mathcal{C}_{anom}  A_0^{m}\binom{n+1}{m+1}+(n-m+1){\tilde C}_{m-1}T_0^{m}}\\
&+  \frac{qe^{-2\sigma}}{\epsilon+p}\left[m \tilde{C}_m T_0^{m+1}-(n+1-m) \tilde{C}_{m-1} A_0  T_0^m\right.\\
&\qquad\left.\qquad+\binom{n+1}{m+1}\mathcal{C}_{anom}A_0^{m+1}\right]
\end{split}
\end{equation}
We want to now use this to obtain the transport coefficients $\xi_m$  in the last relation of \eqref{conscor}.
For this we have to invert the relations \eqref{eq:jxi} for $\xi_m$. We finally get 
\begin{equation}\label{explicitform2n}
\begin{split}
\xi_m &= \brk{m \frac{q\mu}{\epsilon + p}-(m+1)}\mathcal{C}_{anom}\binom{n+1}{m+1} \mu^m\\
&\qquad+\sum_{k=0}^{m}\brk{m \frac{q\mu}{\epsilon + p}-(m-k)}(-1)^{k-1}{\tilde C}_{k}\binom{n-k}{m-k}  T^{k+1}\mu^{m-k-1} \\
\end{split}
\end{equation}
This then is the prediction of this transport coefficient via partition function methods. This 
exactly matches with the expression from \cite{Loganayagam:2011mu} in \eqref{eq:xiLoga} provided we 
make the following identification among the constants $\tilde{C}_m= (-1)^{m-1} C_m$.

\section{Comments on Most Generic Entropy Current}\label{sec:entropy}

Another physical requirement which has long been used as a source of constraints
on fluid dynamical transport coefficients is the local form of second law of 
thermodynamics. As we reviewed in the subsection\S\S\ref{subsec:LogaReview} this principle
had been used in \cite{Loganayagam:2011mu}  to obtain anomaly induced transports 
coefficients in arbitrary even dimensions.\\

In this section we will determine the entropy current in equilibrium by comparing the total 
entropy with that obtained from the equilibrium partition function. In the examples studied in 
\cite{Banerjee:2012iz,Jain:2012rh} it was seen that in general the comparison 
with equilibrium entropy ( obtained from partition function) did not fix 
all the non dissipative coefficients in fluid dynamical entropy current. However
it did determine the anomalous contribution exactly. Here we will see that 
this holds true in general even dimensions. \\ 

Let us begin by computing the entropy from the equilibrium partition function. 
We begin with the anomalous part of the partition function
\begin{equation}\begin{split}
W_{anom} = \frac{1}{T_0}\int d^{2n-1}x  &\sqrt{g_{2n-1}}\bigg\{ \sum_{m=1}^{n} 
         \alpha_{m-1}  \brk{\epsilon A(da)^{m-1}(dA)^{n-m}}\bigg.\\
&\qquad\bigg.\qquad    + \alpha_n \brk{\epsilon a(da)^{n-1}}  \bigg\}
\end{split}
\end{equation}
where the functions $\alpha_m$ are given in \eqref{Csol}.

The anomalous part of the total entropy is easily computed to be 
\begin{equation}\begin{split}\label{entropy1}
S_{anom} &=  \frac{\partial}{\partial T_0} \prn{ T_0 W_{anom} } \\
  &= \int d^{2n-1}x  \sqrt{g_{2n-1}}   \bigg\{ \sum_{m=1}^{n} 
       m~T_0^{m-1} ~\tilde{C}_{m-1}\   \brk{\epsilon A(da)^{m-1}(dA)^{n-m}} \\
  & \quad + ~(n+1) \tilde{C}_n ~T_0^n  \brk{\epsilon a(da)^{n-1}}  \bigg\}\\
  &= \int d^{2n-1}x  \sqrt{g_{2n-1}}    \bigg\{ \sum_{m=1}^{n} 
       (m+1) ~T_0^m ~\tilde{C}_m\ \brk{\epsilon a(da)^{m-1}(dA)^{n-m}} \\
  & \qquad + \tilde{C}_0 \brk{\epsilon A(dA)^{n-1}}  \bigg\}
\end{split}
\end{equation}

Now we will determine the most general form of entropy current in equilibrium by 
comparison with \eqref{entropy1}. In \cite{Banerjee:2012iz} it was argued that the  entropy 
current by itself is not a physical object, but entropy production and total entropy 
are. This gave a window for gauge non invariant contribution to entropy current but 
the contribution was removed by CPT invariance. Here also we will allow for such 
gauge non invariant terms in the entropy current. The most general form of entropy 
current, allowing for gauge non invariant pieces, is then
\begin{equation}
\begin{split}
J^\mu_S &= s u^\mu -\frac{\mu}{T} J^{\mu}_{odd} + \sum_{m=1}^{n} \chi_m \varepsilon^{\mu\nu\ldots} u_\nu (\partial u)^{m-1} 
(\partial\hat{\mathcal{A}})^{n-m}\\
&\qquad + \zeta \varepsilon^{\mu\nu\ldots} \hat{\mathcal{A}}_\nu (\partial\hat{\mathcal{A}})^{n-1}
\end{split}
 \end{equation}
where $\chi_m$ is a function of $T$ and $\mu$ whereas $\zeta$ is a constant
.
The correction to the local entropy density (i.e., the time component of the entropy current) can be written 
after an integration by parts as   
\begin{equation}\label{0entcur}\begin{split}
\delta J_S^0 = \varepsilon^{0ij\ldots}\brk{ \zeta A(dA)^{n-1} + \sum_{k=1}^{n} {\tilde f}_k\ a\ (da)^{k-1}\ (dA)^{n-k} }_{ij\ldots} +\text{total derivatives}
\end{split}
\end{equation}
where
\begin{equation}
\begin{split}
{\tilde f}_m &\equiv  - s U_m +\frac{\mu}{T} J_m+\zeta A_0^m \binom{n}{m}+\sum_{k=1}^{m} \binom{n-k}{m-k}\chi_k  \prn{-e^{\sigma}}^k A_0^{m-k}\\
\end{split}
\end{equation}

The correction to the entropy is then,
\begin{equation}
\begin{split}\label{entropy2}
\delta S &= \int  d^{2n-1}x \sqrt{g_{2n}} ~J^0_S  \\
& = \int  d^{2n-1}x \sqrt{g_{2n-1}} 
     \brk{ \zeta \brk{\epsilon A(dA)^{n-1}} + \sum_{m=1}^{n} {\tilde f}_m\ \brk{\epsilon a\ (da)^{m-1}\ (dA)^{m-k}} } 
\end{split}
\end{equation}

Comparing the two expressions of total equilibrium entropy \eqref{entropy1} and
\eqref{entropy2} we find the following expressions of the various coefficients in
the entropy current \eqref{0entcur},
\begin{eqnarray}\label{resultec}
\zeta &=& {\tilde C}_0 \quad\text{and}\quad \tilde{f}_k = (k+1) ~T_0^{k} ~{\tilde C}_{k} {\rm ~~for~~}  0 \leq k \leq n
\end{eqnarray}

This in turn implies that
\begin{equation}
\begin{split}
T_0\sum_{k=1}^{m} &\binom{n-k}{m-k}\chi_k  \prn{-e^{\sigma}}^k A_0^{m-k}\\
&= \tilde{C}_mT_0^{m+1}+m\binom{n}{m}\mathcal{C}_{anom}A_0^{m+1}-\tilde{C}_0 T_0A_0^m \binom{n}{m}
\end{split}
\end{equation}
which can be inverted to give 
\begin{equation}
\begin{split}
\chi_m &= - \mathcal{C}_{anom}\binom{n+1}{m+1} T^{-1}\mu^{m+1}-\sum_{k=0}^{m}\tilde{C}_k (-1)^{k-1}\binom{n-k}{m-k}  T^k\mu^{m-k} \\
\zeta &= {\tilde C}_0 \\
\end{split}
\end{equation}
which matches with the prediction from \cite{Loganayagam:2011mu} in equation \eqref{eq:Chi_mPrediction}
again with the identification $C_m(-1)^{m-1} = \tilde{C}_m$. We see that in the entropy current 
we have a total of $n+1$ constants as in the equilibrium partition function.

This completes our partition function analysis and our re derivation of the results of \cite{Loganayagam:2011mu}
via partition function techniques. We see that the transport coefficients match exactly with the results
obtained via entropy current (provided the analysis of \cite{Loganayagam:2011mu} is extended by allowing
gauge-non-invariant pieces in the entropy current). This detailed match of transport coefficients
warrants the question whether the form of the equilibrium partition function itself 
can be directly derived  from the expressions of \cite{Loganayagam:2011mu}
quoted in \ref{subsec:LogaReview}. We turn to this question in the next section.

\section{Gibbs current and Partition function}\label{sec:IntByParts}
We begin by repeating the expression for the Gibbs current in \eqref{eq:GCovBOmega}  which was central to the results of 
\cite{Loganayagam:2011mu}.
\begin{equation}
\begin{split}
\bar{\mathcal{G}}^{Cov}_{anom} 
&= C_0 T \hat{\mathcal{A}}\wedge\mathcal{F}^{n-1}+ \sum_{m=1}^{n}\left[\mathcal{C}_{anom}\binom{n+1}{m+1}\mu^{m+1}\right.\\
&\qquad  \left. + \sum_{k=0}^{m}C_k \binom{n-k}{m-k}  T^{k+1}\mu^{m-k}\right] (2\omega)^{m-1} \mathcal{B}^{n-m}\wedge u \\
\end{split}
\end{equation}
The subscript `anom' denotes that we are considering only a part of the entropy current relevant to anomalies.
The superscript `Cov' refers to the fact that this is the Gibbs free energy computed by turning on a chemical
potential for the \textbf{covariant} charge.

Let us ask how this expression would be modified if the Gibbs free energy was computed by turning on a 
chemical potential for the \textbf{consistent} charge instead. The change from covariant charge to 
consistent charge/current is simply given by a shift as  given by the equation\eqref{eq:shift}. 
This shift does not depend on the state of the theory but is purely a functional of the 
background gauge fields. Thinking of Gibbs free energy as minus temperature 
times the logarithm of the Eucidean path integral, a conversion from covariant
charge to a consistent charge induces a shift
\[ \bar{\mathcal{G}}^{Cov}_{anom} = \bar{\mathcal{G}}^{Consistent}_{anom} - \mu\ n\ \mathcal{C}_{anom}\hat{\mathcal{A}}\wedge \mathcal{F}^{n-1}\]
which gives
\begin{equation}\label{eq:GConsBOmega}
\begin{split}
&\bar{\mathcal{G}}^{Consistent}_{anom} \\
&= \sum_{m=1}^{n}\left[\mathcal{C}_{anom}\binom{n+1}{m+1}\mu^{m+1} +\sum_{k=0}^{m}C_k \binom{n-k}{m-k}  T^{k+1}\mu^{m-k}\right] 
(2\omega)^{m-1} \mathcal{B}^{n-m}\wedge u \\
&\qquad + \brk{C_0 T + n\mathcal{C}_{anom}\mu } \hat{\mathcal{A}}\wedge\mathcal{F}^{n-1}\\
\end{split}
\end{equation}
This now a Gibbs current whose $\mu$ derivative gives the consistent current rather than a covariant current. It is 
easy to check that this solves an adiabaticity equation very similar to the one quoted in equation\eqref{eq:adiabG}
\begin{equation}\label{eq:adiabGCons}
\begin{split}
d\bar{\mathcal{G}}^{Consistent}_{anom} &+ \mathfrak{a} \wedge \bar{\mathcal{G}}^{Consistent}_{anom}+n\mathcal{C}_{anom}\prn{\hat{\mathcal{A}}+\mu u}\wedge\mathcal{E}\wedge \mathcal{B}^{n-1}\\
&= \prn{dT+\mathfrak{a}T}\wedge \frac{\partial\bar{\mathcal{G}}^{Consistent}_{anom}}{\partial T}
+ \prn{d\mu+\mathfrak{a}\mu-\mathcal{E}}\wedge \frac{\partial\bar{\mathcal{G}}^{Consistent}_{anom}}{\partial \mu}
\end{split}
\end{equation}
The question we wanted to address is how this Gibbs current is related to the partition function in 
equation \eqref{action}.

The answer turns out to be quite intuitive - we would like to argue in this section that
\begin{equation}\label{eq:ZGibbs}
W_{anom} = \ln\ Z^{anom}_{Consistent} = - \int_{space}\frac{1}{T} \bar{\mathcal{G}}^{Consistent}_{anom}
\end{equation}
This equation instructs us to pull back the $2n-1$ form in equation \eqref{eq:GConsBOmega} (divided by local temperature)
and integrate it on an arbitrary spatial hyperslice to obtain the anomalous contribution to 
negative logarithm of the equilibrium path integral. Note that pulling back the Hodge dual of
Gibbs current on a spatial hyperslice is essentially equivalent to integrating its zero
component (i.e., the Gibbs density) on the slice. Seen this way the above relation is 
the familiar statement relating Gibbs free energy to the grand-canonical partition function.

\subsection{Reproducing the Gauge variation}
Before giving an explicit proof of the relation\eqref{eq:ZGibbs} we will check in this
subsection that the relation\eqref{eq:ZGibbs} essentially gives the correct gauge variation to the path-integral
at equilibrium. This will provide us with a clearer insight on how the program of \cite{Banerjee:2012iz} to
write a local expression in the partition function to reproduce the anomaly works. 

The gauge variation of\eqref{eq:ZGibbs} under $\delta\hat{\mathcal{A}}=d\delta\lambda$ is
\begin{equation}
\begin{split}
\delta W_{anom} &= \delta \ln\ Z^{anom}_{Consistent} = - \int_{space}\frac{1}{T} \delta\bar{\mathcal{G}}^{Consistent}_{anom}\\
&= -\int_{space}\brk{C_0  + n\mathcal{C}_{anom}\frac{\mu}{T} } \delta\hat{\mathcal{A}}\wedge\mathcal{F}^{n-1}\\
&= -\int_{space}\brk{C_0  + n\mathcal{C}_{anom}\frac{\mu}{T} } d\delta\lambda\wedge\mathcal{F}^{n-1}\\
&= -\int_{surface}\delta\lambda\brk{C_0  + n\mathcal{C}_{anom}\frac{\mu}{T} } \wedge\mathcal{F}^{n-1}
 + n\mathcal{C}_{anom}\int_{space}\delta\lambda d\prn{\frac{\mu}{T}} \wedge\mathcal{F}^{n-1}
\end{split}
\end{equation}

We will now ignore the surface contribution and use the fact that chemical equilibrium demands that 
\[ Td\prn{\frac{\mu}{T}} = \mathcal{E} \]
where  $\mathcal{E}\equiv u^\nu\mathcal{F}_{\mu\nu}dx^\nu$ is the rest frame electric-field. This is
essentially a statement (familiar from say semiconductor physics) that in equilibrium the 
diffusion current due to concentration gradients should cancel the drift ohmic current due to the
electric field. Putting this in along with the electric-magnetic decomposition 
$\mathcal{F}=\mathcal{B}+u\wedge \mathcal{E}$, we get  
\begin{equation}
\begin{split}
\delta W_{anom} &= \delta \ln\ Z^{anom}_{Consistent} = \mathcal{C}_{anom}\int_{space}\frac{\delta\lambda}{T} n\mathcal{E}\wedge\mathcal{B}^{n-1}
\end{split}
\end{equation}
which is the correct anomalous variation required of the equilibrium path-integral !
In $d=2n=4$ dimensions for example we get the correct $E.B$ variation along with the 
$1/T$ factor coming from the integration over euclidean time-circle. The factor of $n$
comes from converting to electric and magnetic fields 
\[ \mathcal{F}^n = n\ u\wedge \mathcal{E}\wedge\mathcal{B}^{n-1} \]
Thus the shift piece along with the chemical equilibrium conspires to reproduce the 
correct gauge variation. The reader might wonder why this trick cannot be made to work
by just keeping the shift term alone in the Gibbs current - the  answer is of course
that other terms are required if one insists on adiabaticity in the sense that
we want to solve \eqref{eq:adiabGCons}.

\subsection{Integration by parts}
In this subsection we will prove \eqref{eq:ZGibbs} explicitly. We will begin by evaluating the 
consistent Gibbs current in the equilibrium configuration. We will as before
work in the `zero $\mu_0$' gauge. 

Using the relations in the  appendix~\ref{app:hydrostatics} we get the consistent Gibbs current as  
\begin{equation}
\begin{split}
-\frac{1}{T}&\bar{\mathcal{G}}^{Consistent}_{anom}\\
&=\frac{1}{T_0}\sum_{m=1}^{n}\left[ C_m(-1)^{m-1} T_0^{m+1}-C_0(-1)^{0-1}\binom{n}{m}T_0A_0^m\right.\\
&\qquad\left. -\binom{n}{m+1}\mathcal{C}_{anom}A_0^{m+1}\right] (da)^{m-1}(dA)^{n-m}\wedge (dt+a)\\
&\quad -\frac{1}{T_0} \brk{n \mathcal{C}_{anom}A_0 +C_0T_0}  A\wedge (dA+A_0 da)^{n-1}\\
&\quad -\frac{(n-1)}{T_0} \brk{n \mathcal{C}_{anom}A_0 +C_0T_0} A\wedge dA_0\wedge(dt+a)\wedge (dA+A_0 da)^{n-2}\\
\end{split}
\end{equation}

After somewhat long set of manipulations one arrives at the following form for the consistent Gibbs current
\begin{equation}
\begin{split}
-\frac{1}{T}&\bar{\mathcal{G}}^{Consistent}_{anom}\\
&=  d\left\{\frac{A}{T_0} 
\sum_{m=1}^{n-1}\left[ C_m(-1)^{m-1} T_0^{m+1}-C_0(-1)^{0-1}\binom{n-1}{m}T_0A_0^m\right.\right.\\
&\qquad\left.\left. +m\binom{n}{m+1}\mathcal{C}_{anom}A_0^{m+1}\right] (da)^{m-1}(dA)^{n-1-m}\wedge (dt+a)\right\}\\
&\quad +\frac{A}{T_0}\sum_{m=1}^{n}\brk{C_{m-1}(-1)^{m-2} T_0^m-\binom{n}{m} \mathcal{C}_{anom} A_0^m}(da)^{m-1}(dA)^{n-m}\\
&\quad+ C_n(-1)^{n-1} T_0^{n}(da)^{n-1}\wedge (dt+a)\\
\end{split}
\end{equation}
Here we have taken out a surface contribution which we will suppress from now on since it does not contribute
to the partition function. This final form is easily checked term by term and we will leave that as an exercise
to the reader.

Suppressing the surface contribution we can write
\begin{equation}
\begin{split}
-\frac{1}{T}&\bar{\mathcal{G}}^{Consistent}_{anom}\\
&=  d\brk{\ldots} +\frac{A}{T_0}\sum_{m=1}^{n}\brk{C_{m-1}(-1)^{m-2} T_0^m-\binom{n}{m} \mathcal{C}_{anom} A_0^m}(da)^{m-1}(dA)^{n-m}\\
&\quad+ C_n(-1)^{n-1} T_0^{n}(da)^{n-1}\wedge (dt+a)\\
&=d\brk{\ldots} +\frac{A}{T_0}\wedge 
\sum_{m=1}^{n}\alpha_{m-1}(da)^{m-1}(dA)^{n-m} + \frac{dt+a}{T_0} \wedge \alpha_n(da)^{n-1}  \\
\end{split}
\end{equation}
where we have defined 
\begin{equation}\label{eq:alphaC}
\begin{split}
 \alpha_m &= C_{m}(-1)^{m-1} T_0^{m+1}-\binom{n}{m+1} \mathcal{C}_{anom} A_0^{m+1}\quad \text{for}\ m<n\\
 \alpha_n &= C_{n}(-1)^{n-1} T_0^{n+1}\\
\end{split}
\end{equation}

To get the contribution to the equilibrium partition function, 
we integrate the above equation over the spatial slice (putting $dt=0$). 
We will neglect surface contributions  to get
\begin{equation}
\begin{split}
&\prn{\ln \mathcal{Z}}^{Consistent}_{anom} \\
&=\int_{\text{space}}\frac{A}{T_0}\wedge 
\sum_{m=1}^{n}\brk{C_{m-1}(-1)^{m-2} T_0^m-\binom{n}{m} \mathcal{C}_{anom} A_0^m}(da)^{m-1}(dA)^{n-m} \\
&\qquad + \int_{\text{space}}C_n(-1)^{n-1} T_0^n a \wedge(da)^{n-1}  \\
&=\int_{\text{space}}\frac{A}{T_0}\wedge 
\sum_{m=1}^{n}\alpha_{m-1}(da)^{m-1}(dA)^{n-m} + \int_{\text{space}} \frac{a}{T_0} \wedge \alpha_n(da)^{n-1}  \\
\end{split}
\end{equation}
with $\alpha_{m}$s given by \eqref{eq:alphaC}. We are essentially done - we have got the form in 
\eqref{action} and comparing the equations \eqref{eq:alphaC} and \eqref{Csol} we find a perfect
agreement with the usual relation $C_m(-1)^{m-1}=\tilde{C}_m$. Now by varying this 
partition function we can obtain currents as before (the variation can be directly done in 
form language using the equations we provide in appendix~\ref{app:variationForms}).
With this we have completed a whole circle showing that the two formalisms for anomalous transport 
developed in \cite{Loganayagam:2011mu} and \cite{Banerjee:2012iz} are 
completely equivalent.

Before we conclude, let us rewrite the partition function in terms of the polynomial 
$\mathfrak{F}^\omega_{anom}[T,\mu]$ as  
\begin{equation}
\begin{split}
&\prn{\ln \mathcal{Z}}^{Consistent}_{anom} \\
&=\int_{\text{space}}\frac{A}{T_0 da}\wedge 
\brk{ \frac{ \mathfrak{F}^\omega_{anom}[-T_0 da, dA]-\mathfrak{F}^\omega_{anom}[-T_0 da, 0]}{dA}
-\frac{ \mathfrak{F}^\omega_{anom}[0, dA+A_0 da]}{dA+A_0 da}}\\
&\qquad + \int_{\text{space}}\frac{\mathfrak{F}^\omega_{anom}[-T_0 da, 0]}{(T_0da)^2}\wedge T_0 a \\
\end{split}
\end{equation}
We will consider an example. Using adiabaticity arguments, the authors of \cite{Loganayagam:2012pz}
derived the following expression for
a theory of free Weyl fermions in $d=2n$ spacetime dimensions
\begin{equation}
\begin{split}
\prn{{\mathfrak{F}}^\omega_{anom}}^{free\ Weyl}_{d=2n}&=- 2\pi\sum_{species}  \chi_{_{d=2n}} \brk{\frac{\frac{\tau}{2}T}{\sin \frac{\tau}{2}T}e^{\frac{\tau}{2\pi}q\mu}}_{\tau^{n+1}} \\
\end{split}
\end{equation}
where $\chi_{_{d=2n}}$ is the chirality and the subscript $\tau^{n+1}$ denotes that one needs to Taylor-expand in $\tau$ and
retain the coefficient of $\tau^{n+1}$. Substituting this into the above expression gives the anomalous part 
of the partition function of free Weyl fermions.

\section{Fluids charged under multiple $U(1)$ fields}\label{sec:2ndimmul}

In this section, we will generalize our results to cases where 
we have multiple abelian $U(1)$ gauge fields in arbitrary $2n-$dimensions.

We can take 
\begin{equation}\label{eq:FOmegaCmulti}
\begin{split}
\mathfrak{F}^\omega_{anom}[T,\mu] &= \mathcal{C}_{anom}^{A_1 \ldots A_{n+1}}\mu_{A_1} \ldots \mu_{A_{n+1}}+\sum_{m=0}^{n}C_m^{A_1\ldots A_{n-m}} T^{m+1}\mu_{A_1\ldots A_{n-m}}.\\
\end{split}
\end{equation}
In this case,
the anomaly equation takes the following form,

\begin{equation}\label{anomeq}
\nabla_{\mu}  J^{\mu,A_{n+1}}_{Cov} =\frac{n+1}{2^n} {\cal C}_{anom}^{A_1 A_2\ldots A_{n+1}} \varepsilon^{\mu_1\nu_1\mu_2\nu_2 \ldots \mu_n\nu_n}
\prn{\mathcal{F}_{\mu_1 \nu_1}}_{A_1} \ldots \prn{\mathcal{F}_{\mu_n \nu_n}}_{A_n} .
\end{equation}
Where, in $2n$ dimensions $\can$ has $n+1$ indices denoted by $(A_1,A_2 \cdot A_{n+1})$ and it is symmetric in all its indices.
It is  straightforward to carry on the above computation for the case of multiple $U(1)$
charges and most of the computations remains the same.
Now, for the multiple $U(1)$ case, in partition function \ref{action} the functions $\alpha_m$ and the
constants $\tilde{C}_m$ (and the constants $C_m$ appearing in $\mathfrak{F}^\omega_{anom}$)
have $n-m$ number of indices which are contracted with $n-1-m$ 
number of $dA$ and one $A$. The constant $\zeta$ appearing in the entropy current has 
$n$ indices.

The constant $\tilde{C}_{n}$ (and $\alpha_n$) has no index. All these constants
are symmetric in their indices. Considering the above index structure into account, we can 
understand that the functions $U_m$ appearing in velocity correction and $\chi_m$ appearing in
entropy corrections has $n-m$ indices and the function $J_{m}$ appearing in the charge current
 has $n-m+1$ indices. Now, we can write the generic  form of these functions as follows:
\begin{equation}
\begin{split}
U_m^{A_1A_2\ldots A_{n-m}}&=-\frac{e^{-2\sigma}}{\epsilon+p} \left[ m\tilde C_{m}^{A_1A_2\ldots A_{n-m}}T_0^{m+1}\right.\\
&\qquad  -(n+1-m)\tilde C_{m-1}^{A_1A_2\ldots A_{n-m}B_1}(A_0)_{B_1}T_0^m \\
&\qquad \left.
+\binom{n+1}{m+1}\mathcal{C}_{anom}^{A_{1}\ldots A_{n-m}B_{1}\ldots B_{m+1}}(A_{0})_{B_{1}}\ldots (A_0)_{B_{m+1}}\right]\\
\end{split}
\end{equation}
where $\prn{A_0}_{B_1}$ comes from the $B_1$th gauge field. 
 
Similarly, we can write the coefficients appearing in  $A$'th charge current ($J^{A}$) as,
\begin{equation}\label{transport}
\begin{split}
\prn{J^A}_m^{A_1 A_2\ldots A_{n-m}}&=e^{-\sigma}\left[- (m+1) \mathcal{C}_{anom}^{A A_1 \ldots A_{n-m}B_{1}\ldots B_m}  (A_0)_{B_{1}}\ldots (A_0)_{B_{m}}\binom{n+1}{m+1}\right.\\
&\qquad \left.+ (n-m+1){\tilde C}_{m-1}^{A A_1\ldots A_{n-m}}T_0^m \right]\\
&\qquad +\frac{q^A e^{-2\sigma}}{\epsilon+p} \left[ m\tilde C_{m}^{A_1A_2\ldots A_{n-m}}T_0^{m+1}\right.\\
&\qquad  -(n+1-m)\tilde C_{m-1}^{A_1A_2\ldots A_{n-m}B_1}(A_0)_{B_1}T_0^m \\
&\qquad \left.
+\binom{n+1}{m+1}\mathcal{C}_{anom}^{A_{1}\ldots A_{n-m}B_{1}\ldots B_{m+1}}(A_{0})_{B_{1}}\ldots (A_0)_{B_{m+1}}\right]\\ 
\end{split}
\end{equation}

We can also express the transport coefficients for fluids charged under multiple $U(1)$ charges,
generalising equation \eqref{explicitform2n} as,
\begin{equation}
\begin{split}
&\prn{\xi^A}_m^{A_1 A_2\ldots A_{n-m}} \\
&\ = \brk{m \frac{q^A\mu_B}{\epsilon + p}-(m+1)\delta^A_B}\mathcal{C}_{anom}^{BA_1\ldots A_{n-m}B_1\ldots B_m}
\binom{n+1}{m+1} \mu_{B_1}\ldots\mu_{B_m}\\
&\quad +\sum_{k=0}^{m-1}\brk{m \frac{q^A\mu_B}{\epsilon + p}-(m-k)\delta^A_B}\\
&\qquad \times (-1)^{k-1}{\tilde C}_{k}^{BA_1\ldots A_{n-m}B_1\ldots B_{m-k-1}}\binom{n-k}{m-k}  T^{k+1}\mu_{B_1} \ldots \mu_{B_{m-k-1}} \\
&\quad +\brk{m \frac{q^A}{\epsilon + p}} (-1)^{m-1}{\tilde C}_{m}^{A_1\ldots A_{n-m}}  T^{m+1} \\
\end{split}
\end{equation}

Similarly the coefficieints $\chi_m$ appearing entropy current become
\begin{equation}
\begin{split}
\chi_m^{A_{1}\ldots A_{n-m}} &= - \mathcal{C}_{anom}^{A_{1}\ldots A_{n-m} B_1\ldots B_{m+1}}\binom{n+1}{m+1} T^{-1}\mu_{B_{1}}\ldots \mu_{B_{m+1}}\\
&-\sum_{k=0}^{m}(-1)^{k-1}\binom{n-k}{m-k}  T^k \tilde{C}_k^{A_{1}\ldots  A_{n-m}B_1\ldots B_{m-k}} \mu_{B_{1}}\ldots \mu_{B_{m-k}}\\
\end{split}
\end{equation}

This finishes the analysis of anomalous fluid charged under multiple abelian $U(1)$ gauge fields.

\section{CPT Analysis}\label{sec:CPT}
In this section we analyze the constraints of 2n dimensional CPT invariance on the analysis
of our previous sections.

\begin{table}
\centering
\begin{tabular}[h]{|c|c|c|c|c|}
\hline
Name & Symbol &  CPT \\
\hline
Temperature & $T$ &  + \\
\hline
Chemical Potential & $\mu$ & - \\
\hline
Velocity 1-form & $u$ & + \\
\hline
Gauge field 1-form &$\hat{\mathcal{A}}$ & - \\
\hline
Exterior derivative & $d$ & - \\
\hline
Field strength 2-form & $\mathcal{F}=d\hat{\mathcal{A}}$ & +\\
\hline
Magnetic field 2-form & $\mathcal{B}$ & +\\
\hline
Vorticity 2-form & $\omega$ & -\\
\hline
\end{tabular}
\caption{\label{tab:CPTform}Action of CPT on various forms}
\end{table}

Let us first examine the CPT transformation of the Gibbs current proposed in 
\cite{Loganayagam:2011mu}. Using the Table\S\ref{tab:CPTform}
 we see that the Gibbs current in Eqn.\eqref{eq:GCovBOmega}
is CPT-even provided the coefficients $\{\mathcal{C}_{anom},C_{2k+1}\}$ 
are CPT-even and the coefficients $C_{2k}$ are CPT-odd. Since in a
CPT-invariant theory all CPT-odd coefficients should vanish, we
conclude that $C_m=0$ for even $m$. This conclusion can be phrased
as 
\begin{equation} CPT\quad : \quad C_m(-1)^{m-1}= C_m \end{equation}
Note that this is the same conclusion as reached by assuming the 
relation to the anomaly polynomial.

Next we analyze the constraints of 2n dimensional CPT invariance on the partition function
\eqref{action}.  Our starting point is a partition function of the fluid and we expect it
to be invariant under $2n$dimensional CPT transformation of the fields.
Table\S\ref{cpttab} lists the effect of 2n dimensional C, P and T transformation on various field
appearing in the partition function \eqref{action}. Since $a_i$ is even while $A_i$ and $\partial_j$ are odd under CPT,
the term with coefficient $C_m$ picks up a factor of $(-1)^{(m+1)}$.
Thus CPT invariance tells us that $C_m$ must be
\begin{itemize}
 \item even function of $A_0$ for odd $m$.
 \item odd function of $A_0$ for even $m$.
\end{itemize}
Now the coefficients $C_m$ are fixed upto constants $\tilde{C}_m$ by the requirement that
the partition function reproduces the correct anomaly. Note that the $A_0$(odd under CPT) dependence of
the coefficients $C_m$ thus determined are consistent with the requirement CPT invariance.
Further, CPT invariance forces $\tilde{C}_m = 0$ for even $m$. The last term in the partition function
\eqref{action} is odd under parity and thus its coefficient is set to zero by CPT for
even $n$ whereas for odd $n$ it is left unconstrained.

Thus finally we see that CPT invariance allows for a total of
\begin{itemize}
 \item $\frac{n}{2}$ constant ($\tilde{C}_m$ with $m$ odd) for even $n$.
 \item $\frac{n+1}{2}$ constants ($\tilde {C}_m$ with $m$ even and $\tilde {C}_n$) for odd $n$.
\end{itemize}

In particular the coefficient $\tilde C_0$ always vanishes and thus, for a CPT invariant theory, we never get 
the gauge-non invariant contribution to th elocal entropy current.

\begin{table}
\centering
\begin{tabular}[h]{|c|c|c|c|c|}
\hline
fields & C & P & T & CPT \\
\hline
$\sigma$ & + & + & + & + \\
\hline
$a_i$ & + & - & - & + \\
\hline
$g_{ij}$ & + & + & + & + \\
\hline
$A_0$ & - & + & + & - \\
\hline
$A_i$ & - & - & - & - \\
\hline
\end{tabular}
\caption{\label{cpttab} Action of CPT on various field}
\end{table}

\section{Conclusion}\label{sec:conclusion}
In this paper we have shown that the results of \cite{Kharzeev:2011ds, Loganayagam:2011mu} can
based on entropy arguments can be re derived within a more field-theory friendly  
partition function technique\cite{Banerjee:2012iz,Jain:2012rh,Jensen:2012jh,Jensen:2012jy}.
This has led us to a deeper understanding linking the local description of anomalous
transport in terms of a Gibbs current \cite{Loganayagam:2011mu,Loganayagam:2012pz} to
the global description in terms of partition functions.

An especially satisfying result is that the polynomial structure of anomalous transport 
coefficients discovered in \cite{Loganayagam:2011mu} is reproduced at the level of 
partition functions. There it was shown that the whole set of anomalous transport 
coefficients are  essentially governed by a single homogeneous polynomial
$\mathfrak{F}^\omega_{anom}[T,\mu]$ of temperature and chemical potentials.
The authors of \cite{Loganayagam:2012pz} noticed that in  a free theory of
chiral fermions this polynomial structure is directly linked to the corresponding
anomaly polynomial of chiral fermions via a replacement rule
\begin{equation}
\begin{split}
\mathfrak{F}_{anom}^\omega[T,\mu] = \mathcal{P}_{anom} \brk{ \mathcal{F}  \mapsto \mu, p_1(\mathfrak{R}) \mapsto - T^2 , p_{k>1}(\mathfrak{R}) \mapsto 0 }
\end{split}
\end{equation}
This result could be generalised for an arbitrary free theory with chiral
fermions and chiral p-form fields using sphere partition
function techniques  which link this polynomial
to a specific thermal observable\cite{futureLoga}. 

Various other known results (for example in AdS/CFT) 
support the conjecture that this rule is probably true in all 
theories with some mild assumptions. While we
have succeeded in reproducing the polynomial structure we have not tried
in this paper to check the above conjecture - this necessarily involves a
similar analysis keeping track of the effect of gravitational anomalies which
we have ignored in our work. It would be interesting to extend our analysis
to theories with gravitational anomalies\footnote{As we were finalising this manuscript, a
paper\cite{Valle:2012em} dealing with $1+1$d gravitational anomalies appeared in arXiv.
We thank Amos Yarom for various discussions regarding this topic.}.

We have derived in this paper a particular contribution to the equilibrium partition function
that is linked to the underlying anomalies of the theory. A direct test of this result would
be to do a direct holographic computation of the same quantity in AdS/CFT to obtain these
contributions. Since the CFT anomalies are linked to the Chern-Simons terms in the bulk
the holographic test would be a computation of a generalised Wald entropy for a black hole
solution of a gravity theory with Chern-Simons terms. The usual Wald entropy gets modified
in the presence of such Chern-Simons terms\cite{Tachikawa:2006sz,Bonora:2011gz}  which are
usually a part of higher derivative corrections to gravity. We hope that reproducing 
the results of this paper would give us a test of generalised Wald formalism 
for such higher derivative corrections.

We have directly linked the description in terms of a Gibbs current\cite{Loganayagam:2011mu,Loganayagam:2012pz}
satisfying a kind of adiabticity equation to the global description in terms of partition functions.
Further we have noticed in \eqref{eq:GibbsChi} that at least in the case of anomalous transport this Gibbs current 
is closely linked to what has been called `the non-canonical part of the entropy current '
in various entropy arguments\cite{Bhattacharyya:2012ex}. It would be interesting to see whether
this construction can be generalised beyond the anomalous transport coefficients 
to other partition function computations which appear in \cite{Banerjee:2012iz,Jensen:2012jh}.
This would give us a more local interpretation of the various terms appearing in the
partition function linking them to a specific Gibbs free energy transport process. Hence with such 
a result one could directly identify the coefficients appearing in the partition function
as the transport coefficients of the Gibbs current.

Another interesting observation of \cite{Loganayagam:2011mu} apart from the polynomial structure
is that the anomalous transport satisfies an interesting reciprocity type relation \eqref{eq:reciprocity}
- the susceptibility describing the change in the anomalous charge current 
with a small change in vorticity is equal to the susceptibility 
describing the change in the anomalous energy current with a small change 
in magnetic field. While we see that the results of our paper are consistent
with this observation made in \cite{Loganayagam:2011mu}, we have not succeeded
in deriving this relation directly from the partition function. It would be 
interesting to derive such a relation from the partition function hence 
clarifying how such a relation arises in a microscopic description .

Finally as we have emphasised in the introductions one would hope that 
the results of our paper serve as a starting point for  generalising 
the analysis of anomalies to non-equilibrium phenomena. Can one
write down a Schwinger-Keldysh functional which transforms appropriately - 
does this provide new constraints on the dissipative transport coefficients ?
We leave such questions to future work.

\subsection*{Acknowledgements}
We would like to thank Sayantani Bhattacharyya for collaboration in
the initial stages of this project. 
It is a pleasure to thank  Jyotirmoy Bhattacharya, Dileep Jatkar,
Shiraz Minwalla, Mukund Rangamani, Piotr Surowka, Amos Yarom and 
Cumrun Vafa for various useful discussions on ideas presented in this 
paper. Research of NB is supported by NWO Veni grant, The Netherlands. 
RL would like to thank  \textbf{ICTS discussion meeting on 
current topics of research in string theory}
at the International Centre for Theoretical Sciences(TIFR) , IISc Bangalore for 
their hospitality while this work was being completed. RL is supported by the Harvard Society of Fellows 
through a junior fellowship. Finally, RL would  like to thank various colleagues at the 
Harvard society for interesting discussions. Finally, we would like to thank people of India for their generous support
to research in science.

\newpage

\appendix
   \begin{center}
      {\bf APPENDICES}
    \end{center}

\section{Results of $(3+1)-$ dimensional and $(1+1)-$ dimensional fluid}\label{app:oldresult}
In this appendix we want to specialise our results to $1+1$ and $3+1$ dimensional
anomalous fluids.By considering local entropy production of the system, the results
for $(3+1)-$ dimensional anomalous fluid were  obtained in \cite{Son:2009tf},
\cite{Neiman:2010zi,Bhattacharya:2011tra}  and
for $(1+1)-$dimensional fluid were obtained in \cite{Dubovsky:2011sk}. The same results have also been
obtained in \cite{Banerjee:2012iz} and \cite{Jain:2012rh} for $(3+1)-$ dimensional and
$(1+1)-$dimensional anomalous fluid respectively, by writing the equilibrium partition
function, the technique that we have followed in this paper. Our goal in this section
is to check that the arbitrary dimension results reduce correctly to these special cases.

\subsection{$(3+1)-$ dimensional anomalous fluids}

Let us consider fluid living in $(3+1)-$dimension and is charged under a $U(1)$ current. Take
\begin{equation}
\begin{split}
\mathfrak{F}^\omega_{anom}[T,\mu] &= \mathcal{C}^{d=4}_{anom} \mu^3+C^{d=4}_0 T\mu^2+C^{d=4}_1 T^2\mu+C^{d=4}_2 T^2\mu\\
\end{split}
\end{equation}
the constants $\{C^{d=4}_0,C^{d=4}_2\}$ if non-zero violate CPT since their subscript indices are even. 

By the replacement rule of \cite{Loganayagam:2012pz} this corresponds
to a theory with the anomaly polynomial 
\begin{equation}
\begin{split}
\mathcal{P}_{anom} &= \mathcal{C}^{d=4}_{anom}\mathcal{F}^3-C^{d=4}_1\ p_{_1}\prn{\mathfrak{R}}\wedge \mathcal{F}\\
\end{split}
\end{equation}
where $p_{_1}\prn{\mathfrak{R}}$ is the first-pontryagin 4-form of curvature.

We have
\[ d\bar{J}_{Consistent} = \mathcal{C}^{d=4}_{anom}\mathcal{F}^2\]
\[ d\bar{J}_{Cov} =  3\mathcal{C}^{d=4}_{anom}\mathcal{F}^2 \]
and their difference is given by
\[ \bar{J}_{Cov} = \bar{J}_{Consistent}+2  \mathcal{C}^{d=4}_{anom}\hat{\mathcal{A}}\wedge \mathcal{F} \]
In components we have
\begin{equation}\label{anomeq4d}
\begin{split}
\nabla_{\mu}  J^{\mu}_{Consistent} &= {\cal C}^{d=4}_{anom} \frac{1}{4}\varepsilon^{\mu \nu \rho \sigma}
\mathcal{F}_{\mu \nu} \mathcal{F}_{\rho \sigma} ,\\
\nabla_{\mu}  J^{\mu}_{Cov} &= 3{\cal C}^{d=4}_{anom} \frac{1}{4}\varepsilon^{\mu \nu \rho \sigma}
\mathcal{F}_{\mu \nu} \mathcal{F}_{\rho \sigma} ,\\
 J^{\mu}_{Cov} &= J^{\mu}_{Consistent} + 2  \mathcal{C}^{d=4}_{anom} \frac{1}{2}\varepsilon^{\mu \nu \rho \sigma}
\hat{\mathcal{A}}_{\nu} \mathcal{F}_{\rho \sigma}
\end{split}
\end{equation}
The anomaly-induced transport coefficients (in Landau frame) in this case are given by 
\begin{equation}\label{eq:xi4d}
\begin{split}
J^{\mu,anom}_{Cov} &= \xi_1^{d=4} \varepsilon^{\mu \nu \rho \sigma}u_\nu\partial_\rho \hat{\mathcal{A}}_{\sigma}
+\xi_2^{d=4} \varepsilon^{\mu \nu \rho \sigma}u_\nu\partial_\rho u_{\sigma}\\
\xi_1^{d=4} &= 3\mathcal{C}^{d=4}_{anom} \mu\brk{\frac{q\mu}{\epsilon + p}-2}
+2C^{d=4}_0   T\brk{\frac{q\mu}{\epsilon + p}-1} +C^{d=4}_1 T^2\mu^{-1}\brk{\frac{q\mu}{\epsilon + p}} \\
\xi_2^{d=4} &= \mathcal{C}^{d=4}_{anom} \mu^2\brk{2\frac{q\mu}{\epsilon + p}-3}+C^{d=4}_0 T\mu\brk{2\frac{q\mu}{\epsilon + p}-2}\\
&\quad +C^{d=4}_1   T^2\mu\brk{2\frac{q\mu}{\epsilon + p}-1} +C^{d=4}_2 T^3\mu^{-1}\brk{2\frac{q\mu}{\epsilon + p}} \\
\end{split}
\end{equation}
and
\begin{equation}\label{eq:chi4d}
\begin{split}
J^{\mu,anom}_{S} &= -\frac{\mu}{T}J^{\mu,anom}_{Cov}+ \chi_1^{d=4} \varepsilon^{\mu \nu \rho \sigma}u_\nu\partial_\rho \hat{\mathcal{A}}_{\sigma}
+\chi_2^{d=4} \varepsilon^{\mu \nu \rho \sigma}u_\nu\partial_\rho u_{\sigma}
+\zeta^{d=4} \varepsilon^{\mu \nu \rho \sigma}\hat{\mathcal{A}}_\nu\partial_\rho \hat{\mathcal{A}}_{\sigma}\\
\mathcal{G}^{\mu,anom}_{Cov} &= -T \chi_1^{d=4} \varepsilon^{\mu \nu \rho \sigma}u_\nu\partial_\rho \hat{\mathcal{A}}_{\sigma}
-T \chi_2^{d=4} \varepsilon^{\mu \nu \rho \sigma}u_\nu\partial_\rho u_{\sigma}
-T\zeta^{d=4} \varepsilon^{\mu \nu \rho \sigma}\hat{\mathcal{A}}_\nu\partial_\rho \hat{\mathcal{A}}_{\sigma}\\
-\zeta^{d=4} &= C^{d=4}_0 \\ 
-\chi_1^{d=4} &= 3\mathcal{C}^{d=4}_{anom} T^{-1}\mu^2+2C^{d=4}_0 \mu+C^{d=4}_1 T \\
-\chi_2^{d=4} &= \mathcal{C}^{d=4}_{anom} T^{-1}\mu^3+C^{d=4}_0 \mu^2  + C^{d=4}_1 T\mu +C^{d=4}_2 T^2 \\
\end{split}
\end{equation}
The anomalous part of the consistent partition function is given by 
\begin{equation}\label{eq:Z4d}
\begin{split}
&\prn{\ln \mathcal{Z}}^{Consistent}_{anom} \\
&=\int_{\text{space}}\frac{A}{T_0}\wedge\left\{ \brk{C^{d=4}_0(-1)T_0- 2\mathcal{C}^{d=4}_{anom} A_0}(dA) 
+\brk{C^{d=4}_1 T_0^2- \mathcal{C}^{d=4}_{anom} A_0^2}(da) \right\}\\
&\qquad + \int_{\text{space}}C^{d=4}_2(-1) T_0^2 a \wedge(da)  \\
&=-\frac{\mathcal{C}^{d=4}_{anom}}{T_0}\int d^3x\sqrt{g_3}\epsilon^{ijk}\brk{ 2A_0 A_i\partial_jA_k + A_0^2A_i\partial_ja_k}\\
&\qquad  - C^{d=4}_0 \int d^3x\sqrt{g_3} \epsilon^{ijk}A_i\partial_jA_k+C^{d=4}_1 T_0\int d^3x\sqrt{g_3}\epsilon^{ijk}A_i\partial_ja_k\\
&\qquad- C^{d=4}_2 T_0^2\int d^3x\sqrt{g_3} \epsilon^{ijk}a_i\partial_ja_k \\
\end{split}
\end{equation}

The results for the equilibrium partition function and the transport coefficients
of the fluid have been obtained in \cite{Banerjee:2012iz} in great detail. We will
now compare the results above against the results there. We begin by first fixing
the relation between the notation here and the notation employed in \cite{Banerjee:2012iz}.
Comparing our partition function in \eqref{eq:Z4d} against Eqn(1.11) of \cite{Banerjee:2012iz}
we get a perfect match with the following relabeling of constants\footnote{We warn the
reader that the wedge notation in \cite{Banerjee:2012iz} differs from the one we use by
numerical factors. So the comparisons are to be made \emph{after} converting to explicit
components to avoid confusion.}
\begin{equation}\label{eq:constantConv4d}
C^{d=4}_{anom} = \frac{C}{6}\ ,\quad
C^{d=4}_0 = -C_0\ ,\quad
C^{d=4}_1 = C_2\ ,\quad
C^{d=4}_2 = -C_1\ 
\end{equation}
The first of these relations also follows independently from comparing our eqn\eqref{anomeq4d}
against the corresponding equations in \cite{Banerjee:2012iz} for covariant/consistent anomaly
and the Bardeen current. We then proceed to compare the transport coefficients in Eqn(3.12) and Eqn.(3.21) of 
\cite{Banerjee:2012iz} against our results in \eqref{eq:xi4d} and \eqref{eq:chi4d}.

We get a match provided one uses (in addition to \eqref{eq:constantConv4d} )
the following relations arising from comparing definitions here against  \cite{Banerjee:2012iz} 
\begin{equation}
\xi_B=\xi_1^{d=4}\ ,\quad 
\xi_\omega=2\xi_2^{d=4}\ ,\quad 
D_B=\chi_1^{d=4}\ ,\quad 
D_\omega=2\chi_2^{d=4}\ ,\quad 
h= \zeta^{d=4}
\end{equation}

\subsection{$(1+1)-$ dimensional anomalous fluids}
Let us consider fluid living in $(1+1)-$dimension and is charged under a $U(1)$ current. Take
\begin{equation}
\begin{split}
\mathfrak{F}^\omega_{anom}[T,\mu] &= \mathcal{C}^{d=2}_{anom} \mu^2+C^{d=2}_0 T\mu+C^{d=2}_1 T^2\\
\end{split}
\end{equation}
the constant $C^{d=2}_0$ if non-zero violates CPT since its subscript index is even. 

By the replacement rule of \cite{Loganayagam:2012pz} this corresponds
to a theory with the anomaly polynomial 
\begin{equation}
\begin{split}
\mathcal{P}_{anom} &= \mathcal{C}^{d=2}_{anom}\mathcal{F}^2-C^{d=2}_1\ p_{_1}\prn{\mathfrak{R}}\\
\end{split}
\end{equation}
where $p_{_1}\prn{\mathfrak{R}}$ is the first-pontryagin 4-form of curvature.

We have
\[ d\bar{J}_{Consistent} = \mathcal{C}^{d=2}_{anom}\mathcal{F}\]
\[ d\bar{J}_{Cov} =  2\mathcal{C}^{d=2}_{anom}\mathcal{F} \]
and their difference is given by
\[ \bar{J}_{Cov} = \bar{J}_{Consistent}+  \mathcal{C}^{d=2}_{anom}\hat{\mathcal{A}} \]
In components we have
\begin{equation}\label{anomeq2d}
\begin{split}
\nabla_{\mu}  J^{\mu}_{Consistent} &= {\cal C}^{d=2}_{anom} \frac{1}{2}\varepsilon^{\mu \nu }
\mathcal{F}_{\mu \nu} ,\\
\nabla_{\mu}  J^{\mu}_{Cov} &= 2{\cal C}^{d=2}_{anom} \frac{1}{2}\varepsilon^{\mu \nu }
\mathcal{F}_{\mu \nu}  ,\\
 J^{\mu}_{Cov} &= J^{\mu}_{Consistent} +   \mathcal{C}^{d=2}_{anom} \varepsilon^{\mu \nu }
\hat{\mathcal{A}}_{\nu} 
\end{split}
\end{equation}
The anomaly-induced transport coefficients (in Landau frame) in this case are given by 
\begin{equation}
\begin{split}
J^{\mu,anom}_{Cov} &= \xi_1^{d=2} \varepsilon^{\mu \nu }u_\nu\\
\xi_1^{d=2} &= \mathcal{C}^{d=2}_{anom} \mu\brk{\frac{q\mu}{\epsilon + p}-2}+C^{d=2}_0 T\brk{\frac{q\mu}{\epsilon + p}-1}
+C^{d=2}_1   T^2\mu^{-1}\brk{\frac{q\mu}{\epsilon + p}}  \\
\end{split}
\end{equation}
and
\begin{equation}
\begin{split}
J^{\mu,anom}_{S} &= -\frac{\mu}{T}J^{\mu,anom}_{Cov}+ \chi_1^{d=2} \varepsilon^{\mu \nu }u_\nu
+\zeta^{d=2} \varepsilon^{\mu \nu }\hat{\mathcal{A}}_\nu\\
\mathcal{G}^{\mu,anom}_{Cov} &= -T \chi_1^{d=2} \varepsilon^{\mu \nu }u_\nu
-T\zeta^{d=2} \varepsilon^{\mu \nu }\hat{\mathcal{A}}_\nu\\
-\zeta^{d=2} &= C^{d=2}_0 \\ 
-\chi_1^{d=2} &= \mathcal{C}^{d=2}_{anom} T^{-1}\mu^2+C^{d=2}_0 \mu+C^{d=2}_1 T \\
\end{split}
\end{equation}
The anomalous part of the consistent partition function is given by 
\begin{equation}\label{eq:Z2d}
\begin{split}
&\prn{\ln \mathcal{Z}}^{Consistent}_{anom} \\
&=\int_{\text{space}}\frac{A}{T_0}\wedge \brk{C^{d=2}_0(-1)T_0- \mathcal{C}^{d=2}_{anom} A_0} 
+ \int_{\text{space}}C^{d=2}_1 T_0 a  \\
&=-\frac{\mathcal{C}^{d=2}_{anom}}{T_0}\int dx\sqrt{g_1}\epsilon^{i}A_0 A_i - C^{d=2}_0 \int dx\sqrt{g_1}\epsilon^{i}A_i
+C^{d=2}_1 T_0\int dx\sqrt{g_1}\epsilon^{i}a_i\\
\end{split}
\end{equation}

Now we are all set to compare our results with the results of \cite{Jain:2012rh}. The comparison
proceeds here the same way as the comparison in $3+1$d before. By comparing Eqn(2.4) of 
\cite{Jain:2012rh} against our \eqref{eq:Z2d} we get\footnote{Note that 
authors of \cite{Jain:2012rh} set the CPT-violating coefficient $C^{d=2}_0= -C_1=0 $ in
most of their analysis. This fact has to be accounted for during the comparison.}

\begin{equation}\label{eq:constantConv2d}
C^{d=2}_{anom} = C\ ,\quad
C^{d=2}_0 = -C_1 \ ,\quad
C^{d=2}_1 = -C_2\ ,\quad
\end{equation}
and we get a match of transport coefficients using the definitions
\begin{equation}
\xi_j=\xi_1^{d=2}\ ,\quad 
\xi_s+\frac{\mu}{T}\xi_j=\chi_1^{d=2}\ ,\quad 
D_\omega=2\chi_2^{d=4}\ ,\quad 
h= \zeta^{d=2}
\end{equation}

\section{Hydrostatics and Anomalous transport}\label{app:hydrostatics}
In this section we will follow \cite{Banerjee:2012iz,Jain:2012rh} in describing a hydrostatic configuration,i.e.,
a time-independent hydrodynamic configuration in a gauge/gravitational background. We will then 
proceed to evaluate the anomalous currents derived in previous section in this background.
This is followed by a computation of consistent partition function by integrating the
consistent Gibbs current over a spatial slice. For convenience we will phrase 
our entire discussion in the language of forms (as in the 
previous section) and refer the reader to the appendix\ref{app:formConventions} for our
form conventions.

Let us consider the special case where we consider a stationary (time-independent) spacetime 
with a metric given by 
\[ g_{spacetime}= -\gamma^{-2}(dt+a)^2 + g_{space} \]
where in the notation of  \cite{Banerjee:2012iz}we can write $\gamma \equiv e^{-\sigma}$. 
Following the discussion there, consider a time-independent fluid configuration with 
local temperature and chemical potential $T ,\mu $ and 

placed in a time-independent gauge-field background 
\[\hat{\mathcal{A}} = \mathcal{A}_0 dt + \mathcal{A} \]
We first compute 
\begin{equation}
\begin{split}
\mathcal{E}& \equiv \mathcal{F}_{\mu\nu}dx^\mu u^\nu = \gamma \mathcal{F}_{i0}dx^i =\gamma d\mathcal{A}_0 \\
\mathfrak{a} &\equiv u^\mu \nabla_\mu u_\nu dx^\nu =-\gamma^{-1} d\gamma= \gamma d\gamma^{-1}\\
dT +\mathfrak{a}T  &= \gamma d\prn{\gamma^{-1}T } \\
d\mu +\mathfrak{a}\mu -\mathcal{E} &= \gamma d\prn{\gamma^{-1}\mu -\mathcal{A}_0} \\
\end{split}
\end{equation}
If we insist that
\begin{equation}
\begin{split}
dT +\mathfrak{a}T  &=0 \\
d\mu +\mathfrak{a}\mu -\mathcal{E} &=0 \\
\end{split}
\end{equation}
then it follows that the quantities 
\[ T_0\equiv \gamma^{-1}T  \quad\text{and}\quad \mu_0\equiv \gamma^{-1}\mu -\mathcal{A}_0 \]
are constant across space. We can invert this to write
\[ T = \gamma T_0 \quad\text{and}\quad \mu  = \gamma \prn{\mathcal{A}_0+\mu_0}\equiv \gamma A_0 \]
where we have defined $A_0 \equiv \mathcal{A}_0 +\mu_0$.Following \cite{Banerjee:2012iz}we will split the gauge field as 
\[\hat{\mathcal{A}} = \mathcal{A}_0 dt + \mathcal{A} = A_0 (dt+a) + A-\mu_0 dt \]
where $A\equiv \mathcal{A}-A_0\ a $. We are now working in a general gauge - often it is useful to
work in a specific gauge : one gauge we will work on is obtained from this generic gauge by 
performing a gauge transformation to remove the $\mu_0 dt$ piece. We will call this gauge
as the `zero $\mu_0$' gauge. In this gauge the new gauge field is given in terms of the
old gauge field via 
\[ \hat{\mathcal{A}}_{\mu_0=0} \equiv  \hat{\mathcal{A}}+\mu_0 dt \]
We will quote all our consistent currents in this gauge.

We are now ready to calculate various hydrostatic quantities 
\begin{equation}\label{eq:HStatics}
\begin{split}
\mathcal{E}& =\gamma d\mathcal{A}_0 =\gamma dA_0 \\
\mathfrak{a} & =-\gamma^{-1} d\gamma= \gamma d\gamma^{-1}\\
\mathcal{B}&\equiv \mathcal{F}-u\wedge \mathcal{E} = d\brk{A_0 (dt+a) + A-\mu_0 dt}+(dt+a)\wedge dA_0\\
&= dA+A_0 da \\
2\omega &= du+u\wedge \mathfrak{a} =- \gamma^{-1} da \\
2\omega T  &=-T_0 da \\
2\omega\mu  &= -A_0 da \\
\hat{A}+\mu  u &= A-\mu_0 dt\\
\mathcal{B}+2\omega \mu  &= dA \\
\end{split}
\end{equation}

Now let us compute the various anomalous currents in terms of the hydrostatic fields.
Using \eqref{eq:HStatics} we get the Gibbs current as
\begin{equation}
\begin{split}
-&\bar{\mathcal{G}}^{Cov}_{anom} \\
&= \gamma \sum_{m=1}^{n}\left[ C_m(-1)^{m-1} T_0^{m+1}-C_0(-1)^{0-1}\binom{n}{m}T_0A_0^m\right.\\
&\qquad\left. +m\binom{n+1}{m+1}\mathcal{C}_{anom}A_0^{m+1}\right] (da)^{m-1}(dA)^{n-m}\wedge (dt+a)\\
&\qquad -\gamma C_0 T_0 \hat{\mathcal{A}}_{\mu_0=0}\wedge \mathcal{F}^{n-1}
\end{split}
\end{equation}
In the following we will always write the minus signs in the form $C_m(-1)^{m-1}$ so that
once we impose CPT all the minus signs could be dropped.

We can now calculate the charge/entropy/energy currents
\begin{equation}
\begin{split}
\bar{J}^{Cov}_{anom} 
&= \sum_{m=1}^{n}\left[-(n+1-m)C_{m-1}(-1)^{m-2} T_0^m\right.\\
&\qquad\left. +(n+1)\binom{n}{m}\mathcal{C}_{anom} A_0^{m}\right] (da)^{m-1}\wedge(dA)^{n-m}\wedge (dt+a) \\
\end{split}
\end{equation}
\begin{equation}
\begin{split}
\bar{J}^{Cov}_{S,anom} 
&=\sum_{m=1}^{n}\left[(m+1)C_m (-1)^{m-1} T_0^m\right.\\
&\qquad \left.-C_0(-1)^{0-1}\binom{n}{m} A_0^m\right] (da)^{m-1}(dA)^{n-m}\wedge (dt+a) \\
&\qquad -C_0  \hat{\mathcal{A}}_{\mu_0=0}\wedge \mathcal{F}^{n-1}
\end{split}
\end{equation}
and
\begin{equation}
\begin{split}
&\bar{q}^{Cov}_{anom}\\ 
&=   \gamma\sum_{m=1}^{n}\left[m C_m (-1)^{m-1} T_0^{m+1}-(n+1-m) C_{m-1}(-1)^{m-2} T_0^m A_0\right.\\
&\qquad\left. +\binom{n+1}{m+1}\mathcal{C}_{anom}A_0^{m+1}\right](da)^{m-1}(dA)^{n-m}\wedge (dt+a) \\
\end{split}
\end{equation}

We can  go to the Landau frame as before
\begin{equation}
\begin{split}
u^\mu &\mapsto u^\mu - \frac{q^\mu_{anom}}{\epsilon + p} \\ 
J^\mu_{anom} &\mapsto J^\mu_{anom} - q \frac{q^\mu_{anom}}{\epsilon + p} \\
J^\mu_{S,anom} &\mapsto J^\mu_{S,anom} - s \frac{q^\mu_{anom}}{\epsilon + p}\\
q^\mu_{anom} &\mapsto 0\\
\end{split}
\end{equation}
In the Landau frame we can write the corrections to various quatities as 
\begin{equation}
\begin{split}
\delta \bar{u} &\equiv -\gamma^{-1}\sum_{m=1}^n U_m (da)^{m-1}\wedge(dA)^{n-m}\wedge(dt+a)\\
\delta \bar{J}^{Cov}_{anom} &\equiv -\gamma^{-1}\sum_{m=1}^n \prn{J_m+ q\ U_m}(da)^{m-1}\wedge(dA)^{n-m}\wedge(dt+a)\\
\delta \bar{J}^{Cov}_{S,anom} &\equiv -\gamma^{-1}\sum_{m=1}^n \prn{S_m+ s\ U_m}(da)^{m-1}\wedge(dA)^{n-m}\wedge(dt+a)\\
\end{split}
\end{equation}
where 
\begin{equation}
\begin{split}
U_m &=-\frac{\gamma^2}{\epsilon + p}\left[ m C_m (-1)^{m-1}T_0^{m+1}-(n+1-m) C_{m-1}(-1)^{m-2} T_0^m A_0\right.\\
&\qquad\left.+\binom{n+1}{m+1}\mathcal{C}_{anom}A_0^{m+1}\right] \\
J_m+ q\ U_m &= \gamma\brk{(n+1-m)C_{m-1}(-1)^{m-2} T_0^m-(n+1)\binom{n}{m}\mathcal{C}_{anom} A_0^{m}}\\
S_m+ s\ U_m &=\gamma\brk{-(m+1)C_m (-1)^{m-1}T_0^m+C_0(-1)^{0-1}\binom{n}{m} A_0^m}\\
\end{split}
\end{equation}
which matches with expressions from the partition function.

The corresponding consistent currents can be obtained via the relations 
\begin{equation}
\begin{split}
\bar{\mathcal{G}}^{Cov}_{anom} &= \bar{\mathcal{G}}^{Consistent}_{anom} - \mu\ n\ \mathcal{C}_{anom}\hat{\mathcal{A}}\wedge \mathcal{F}^{n-1}\\
\bar{J}^{Cov}_{anom} &= \bar{J}^{Consistent}_{anom} + n\ \mathcal{C}_{anom}\hat{\mathcal{A}}\wedge \mathcal{F}^{n-1} \\
\bar{J}^{Cov}_{S,anom} &= \bar{J}^{Consistent}_{S,anom} \\
\bar{q}^{Cov}_{anom} &= \bar{q}^{Consistent}_{anom}\\
\end{split}
\end{equation}
In particular we have 
\begin{equation}
\begin{split}
-\frac{1}{T}&\bar{\mathcal{G}}^{Consistent}_{anom}\\
&=\frac{1}{T_0}\sum_{m=1}^{n}\left[ C_m(-1)^{m-1} T_0^{m+1}-C_0(-1)^{0-1}\binom{n}{m}T_0A_0^m\right.\\
&\qquad\left. -\binom{n}{m+1}\mathcal{C}_{anom}A_0^{m+1}\right] (da)^{m-1}(dA)^{n-m}\wedge (dt+a)\\
&\quad -\frac{1}{T_0} \brk{n \mathcal{C}_{anom}A_0 +C_0T_0}  A\wedge (dA+A_0 da)^{n-1}\\
&\quad -\frac{(n-1)}{T_0} \brk{n \mathcal{C}_{anom}A_0 +C_0T_0} A\wedge dA_0\wedge(dt+a)\wedge (dA+A_0 da)^{n-2}\\
\end{split}
\end{equation}

\section{Variational formulae in forms}\label{app:variationForms}
The energy current is defined via the relation
\begin{equation}
\begin{split}
q_\mu dx^\mu &\equiv -T_{\mu\nu}u^\mu dx^\nu \\
&= -\gamma T_{00} (dt+a) - \gamma g_{ij} T^i_0 dx^j \\
\end{split} 
\end{equation}
Hence its Hodge dual  is  (See \ref{app:formConventions} for the definition of Hodge dual)
\begin{equation}
\begin{split}
\bar{q} &= \gamma^3 T_{00} d\forall_{d-1} + \gamma T^i_0 (dt+a)\wedge \prn{d\Sigma_{d-2}}_i
\end{split} 
\end{equation}
We take the following  relations\footnote{we remind the reader that $\gamma\equiv e^{-\sigma}$ and
$d\forall_{d-1} = d^{d-1}x \sqrt{-\det\ g_d}$} from Eqn(2.16) of \cite{Banerjee:2012iz} 
\begin{equation}
\begin{split}
\gamma T_{00} d\forall_{d-1} &=  \frac{\delta}{\delta \gamma} \prn{T_0 \ln\ \mathcal{Z}} \\
T^i_0 d\forall_{d-1} &= dx^i \wedge T^j_0 \prn{d\Sigma_{d-2}}_j
= \brk{\frac{\delta}{\delta a_i}-A_0 \frac{\delta}{\delta A_i}} \prn{T_0 \ln\ \mathcal{Z}} \\
\end{split} 
\end{equation}
where the independent variables are $\{\gamma, a,g^{ij},A_0,A,T_0,\mu_0\}$. Converting into forms
\begin{equation}
\begin{split}
\bar{q} &= \brk{\gamma^2 \frac{\delta}{\delta \gamma} + \gamma(dt+a)\wedge\frac{\delta}{\delta a}-\gamma A_0(dt+a)\wedge \frac{\delta}{\delta A}}
\prn{T_0\ln \mathcal{Z}}\\
&=\brk{\gamma^2 \frac{\delta}{\delta \gamma} +\gamma(dt+a)\wedge\frac{\delta}{\delta a}-\mu(dt+a)\wedge \frac{\delta}{\delta A}}\prn{T_0\ln \mathcal{Z}} \\
\end{split} 
\end{equation}

Similarly for the charge current 
\begin{equation}
\begin{split}
-\gamma^2 J_0 d\forall_{d-1} &=  \frac{\delta}{\delta A_0} \prn{T_0 \ln\ \mathcal{Z}} \\
J^i d\forall_{d-1} &= dx^i \wedge J^j \prn{d\Sigma_{d-2}}_j
= \frac{\delta}{\delta A_i} \prn{T_0 \ln\ \mathcal{Z}} \\
\end{split} 
\end{equation}
which implies
\begin{equation}
\begin{split}
\bar{J} &\equiv-\gamma^2 J_0 d\forall_{d-1}- J^i(dt+a)\wedge \prn{d\Sigma_{d-2}}_i\\
&= \brk{\frac{\delta}{\delta A_0} -(dt+a)\wedge\frac{\delta}{\delta A} }\prn{T_0\ln \mathcal{Z}}
\end{split}
\end{equation}

Putting $T_0 \ln \mathcal{Z}= -\gamma^{-1}\bar{\mathcal{G}}$ we can write
\begin{equation}
\begin{split}
\bar{J} &\equiv -\frac{\partial \bar{\mathcal{G}}}{\partial\mu}=-\gamma^{-1} \brk{\frac{\delta}{\delta A_0} -(dt+a)\wedge\frac{\delta}{\delta A} }\bar{\mathcal{G}}\\
\bar{J}_S 
&\equiv -\frac{\partial \bar{\mathcal{G}}}{\partial T}=-\gamma^{-1}\frac{1}{T_0}\brk{ \gamma\frac{\delta }{\delta \gamma}+(dt+a)\wedge\frac{\delta}{\delta a}-A_0\frac{\delta}{\delta A_0}  }\bar{\mathcal{G}}
 \\
\bar{q} 
&= \bar{\mathcal{G}}+T\bar{J}_S +\mu \bar{J} \\ 
\end{split} 
\end{equation}


\section{Convention for Forms}\label{app:formConventions}

The inner product between two 1-forms $J\equiv J_0 (dt+a)+ g_{ij} J^i dx^j$ and $J'\equiv J'_0 (dt+a)+ g_{ij} (J')^i dx^j$ 
is given in terms of the KK-invariant components as
\begin{equation}
\begin{split}
\langle J ,J' \rangle &\equiv -\gamma^2 J_0 J'_0 + g_{ij}J^i  (J')^j
\end{split}
\end{equation}

In general, the exterior derivative of a p-form
\[ A_p \equiv \frac{1}{p!}A_{\mu_1\ldots\mu_p} dx^{\mu_1}\wedge\ldots\wedge dx^{\mu_p} \]
is given by
\begin{equation}
\begin{split}
(dA)_{p+1} &\equiv \frac{1}{p!}\partial_\lambda A_{\mu_1\ldots\mu_p} dx^\lambda\wedge dx^{\mu_1}\wedge\ldots\wedge dx^{\mu_p}\\
&=\frac{1}{(p+1)!}\brk{\partial_{\mu_1} A_{\mu_2\ldots\mu_{p+1}}+\text{cyclic}} dx^{\mu_1}\wedge\ldots\wedge dx^{\mu_{p+1}}
\end{split} 
\end{equation}

The Levi-Civita tensor $\varepsilon^{\mu_1\ldots\mu_d}$ is defined as the completely antisymmetric tensor
with
\[ \varepsilon^{012\ldots (d-1)} = \frac{1}{\sqrt{-\det\ g_d}} = \frac{1}{\gamma^{-1}\sqrt{\det\ g_{d-1}}}  \]
We will also often define the spatial Levi-Civita tensor $\epsilon^{i_1i_2\ldots i_{d-1}}$ such that 
\[ \epsilon^{12\ldots (d-1)} = \frac{1}{\sqrt{\det\ g_{d-1}}}   \]
which is related to its spacetime counterpart via 
\[ \epsilon^{i_1i_2\ldots i_{d-1}} = \gamma^{-1} \varepsilon^{0i_1i_2\ldots i_{d-1}}\]

Let us define the spatial volume $(d-1)$-form as 
\begin{equation}
\begin{split}
d\forall_{d-1} &\equiv \gamma^{-1}\epsilon_{i_1\ldots i_{d-1}} dx^{i_1}\otimes\ldots\otimes dx^{i_{d-1}} \\ 
&= \frac{1}{(d-1)!}\gamma^{-1}\epsilon_{i_1\ldots i_{d-1}} dx^{i_1}\wedge\ldots\wedge dx^{i_{d-1}} \\
&= d^{d-1}x\ \gamma^{-1}\sqrt{\det\ g_{d-1}} \\
&= d^{d-1}x\ \sqrt{-\det\ g_d}
\end{split}
\end{equation}
where $\epsilon_{i_1\ldots i_{d-1}}$ is the spatial Levi-Civita symbol.
The form $d\forall_{d-1}$  transforms like a vector with a lower time-index and hence is KK-invariant.

Define the spatial area $(d-2)$-form as 
\begin{equation}
\begin{split}
\prn{d\Sigma_{d-2}}_j &\equiv \gamma^{-1}\epsilon_{j i_1\ldots i_{d-2}} dx^{i_1}\otimes\ldots\otimes dx^{i_{d-2}} \\ 
&= \frac{1}{(d-2)!}\gamma^{-1}\epsilon_{ji_1\ldots i_{d-2}} dx^{i_1}\wedge\ldots\wedge dx^{i_{d-2}} \\
\end{split}
\end{equation}
This transforms like a vector with a lower time-index and a lower spatial index but is antisymmetric in 
these two indices and  is hence KK-invariant. The area $(d-2)$-form satisfies
\[ dx^i\wedge \prn{d\Sigma_{d-2}}_j =  d\forall_{d-1}\ \delta^i_j \]

The Hodge-dual of a 1-form $J\equiv J_0 (dt+a)+ g_{ij} J^i dx^j$ is defined as
\begin{equation}\label{eq:HodgeDef}
\begin{split}
\bar{J} 
&= -\gamma^2 J_0 d\forall_{d-1}- J^i(dt+a)\wedge \prn{d\Sigma_{d-2}}_i\\
\end{split}
\end{equation}
This is defined such that 
\begin{equation}
\begin{split}
J'\wedge \bar{J} = \langle J',J \rangle (dt+a)\wedge d\forall_{d-1} = \langle J' ,J \rangle dt\wedge d\forall_{d-1} 
\end{split}
\end{equation}
In particular 
\begin{equation}
\begin{split}
d\bar{J} = \prn{\nabla_\mu J^\mu} dt\wedge d\forall_{d-1} 
\end{split}
\end{equation}
One often useful formula is this 
\begin{equation}\label{eq:HodgeDual}
\begin{split}
\bar{J} &= \hat{\mathcal{A}}\wedge(d\hat{\mathcal{A}})^{n-1} \\ 
&\qquad\text{is equivalent to}\\
J^\mu &= \brk{\varepsilon \hat{\mathcal{A}}\ (\partial\hat{\mathcal{A}})^{n-1}}^\mu\\
\end{split}
\end{equation}

Let us take another example which will recur throughout our paper - 
say we are given that the Hodge-dual of a 1-form $J\equiv J_0 (dt+a)+ g_{ij} J^i dx^j$
is 
\[ -\bar{J} = A\wedge(da)^{m-1}(dA)^{n-m}+ A_0(dt+a)\wedge (da)^{m-1}(dA)^{n-m} \]
where $a=a_i dx^i$ and $A=A_i dx^i$ are two arbitrary 1-forms with only spatial components. 

Then we can invert the Hodge-dual using the following statement
\begin{equation}\label{eq:InvertHodge}
\begin{split}
\bar{J} &= -A\wedge(da)^{m-1}(dA)^{n-m} - A_0(dt+a)\wedge (da)^{m-1}(dA)^{n-m}\\ 
&\qquad\text{is equivalent to}\\
J_0 &= \gamma^{-1} \brk{\epsilon A(da)^{m-1}(dA)^{n-m}}\\
J^i &= \gamma A_0 \brk{\epsilon(da)^{m-1}(dA)^{n-m}}^i\\
\end{split}
\end{equation}

\bibliographystyle{JHEP}
\bibliography{anomaly}

\end{document}